\documentclass[lettersize,journal]{IEEEtran}

\ifCLASSINFOpdf

\else

\fi
\usepackage{xcolor}
\usepackage{algorithmic}
\usepackage{array}
\usepackage{stfloats}
\usepackage{cite}
\usepackage{graphicx}
\usepackage{epstopdf}
\graphicspath{{Figures/}{../Figures/}}
\usepackage{gensymb,bm}
\usepackage[T1]{fontenc}
\usepackage[utf8]{inputenc}
\usepackage{pdfpages}
\usepackage{textcomp}
\usepackage[cmex10]{amsmath}
\usepackage{booktabs}
\usepackage{url}
\usepackage{subcaption}
\usepackage{url}
\usepackage{multicol}
\usepackage{multirow}
\usepackage{siunitx}
\usepackage{adjustbox}
\usepackage{acronym}

\begin{document}

\acrodef{APS}[APS]{angular power spectrum}
\newacroplural{APS}{angular power spectra}
\acrodef{PDP}[PDP]{power delay profile}
\acrodef{PL}[PL]{path loss}
\acrodef{RMSDS}[RMSDS]{RMS delay spread}
\acrodef{AS}[AS]{angular spread}
\acrodef{MPC}[MPC]{multipath component}
\acrodef{DoD}[DoD]{direction of departure}
\acrodef{DoA}[DoA]{direction of arrival}
\acrodef{LoS}[LoS]{line-of-sight}
\acrodef{OLoS}[OLoS]{obstructed line-of-sight}
\acrodef{NLoS}[NLoS]{non line-of-sight}
\acrodef{Tx}[Tx]{transmitter}
\acrodef{Rx}[Rx]{receiver}
\acrodef{MIMO}[MIMO]{multiple-input multiple-output}
\acrodef{CDF}[CDF]{cummulative distribution function}
\acrodef{5G}[5G]{fifth-generation}
\acrodef{UWB}[UWB]{ultra-wideband}
\acrodef{OTA}[OTA]{over-the-air}
\acrodef{BS}[BS]{base station}
\acrodef{UE}[UE]{user equipment}

\title{Ultra-Wideband Double-Directional Channel Measurements and Statistical Modeling in Urban Microcellular Environments for the Upper-Midband/FR3}
\author{Naveed A. Abbasi, Kelvin Arana, Siddhant Singh, Atulya Bist, Vikram Vasudevan, Tathagat Pal, \\ Jorge Gomez-Ponce, Young-Han Nam, Charlie Zhang, and Andreas F. Molisch
\thanks{N. A. Abbasi, Kelvin Arana, Siddhant Singh, Atulya Bist, Tathagat Pal, Vikram Vasudevan, Jorge Gomez-Ponce and A. F. Molisch are with the Ming Hsieh Department of Electrical and Computer Engineering, University of Southern California, Los Angeles, CA, USA. J. Gomez-Ponce is also with the ESPOL Polytechnic University, Escuela Superior Politécnica del Litoral, ESPOL, Facultad de Ingenier\'ia en Electricidad y Computaci\'on, Km 30.5 vía Perimetral, P. O. Box 09-01-5863, Guayaquil, Ecuador. Young-Han Nam and Charlie Zhang are with Samsung Research America, Richardson, TX, USA. Corresponding author: Naveed A. Abbasi (nabbasi@usc.edu).}
}
\maketitle

\begin{abstract}
The upper midband, designated as Frequency Range 3 (FR3), is increasingly critical for the next-generation of wireless networks. Channel propagation measurements and their statistical analysis are essential first steps towards designing 5G and 6G systems in this band. This paper presents a comprehensive ultra-wideband (UWB) double-directional channel measurement campaign in a large portion of FR3 (6-14 \SI{}{\giga\hertz}) for urban microcellular environments. We analyze over 25,000 directional power delay profiles and provide key insights into line-of-sight (LoS) and obstructed line-of-sight (OLoS) conditions. This is followed by statistical modeling of path loss, shadowing, delay spread and angular spread. As the first UWB double-directional outdoor measurement campaign in this frequency range, this work offers critical insights for spectrum allocation, channel modeling, and the design of advanced communication systems, paving the way for further exploration of FR3. 
\end{abstract}

\IEEEpeerreviewmaketitle

\section{Introduction}

\subsection{Motivation}


Total mobile data traffic is expected to increase by a factor of $2.5$ from $2024$ to $2030$, to $300$ Exabyte per month \cite{ericssonforecast}. One of the ways to satisfy this increased demand is the use of new spectrum bands for cellular communications. Over the past years, frequency regulators have started to make available portions of the upper midband, i.e., $6-24$ \SI{}{\giga\hertz}, and the worldwide cellular standardization body has defined this as ''Frequency Range 3 (FR3)" to complement the previously defined sub-6 \SI{}{\giga\hertz} (FR 1) and millimeter-wave (FR 3) bands.\footnote{3GPP has defined FR 3 actually as $7.125 - 24.25$ \SI{}{\giga\hertz}, mainly because the frequency regulator in the USA, the Federal Communications Commission (FCC) has reserved the $6-7.125$ \SI{}{\giga\hertz} band for WiFi. However, in other parts of the world, the rules are not as clear, and in particular potential use of this band for unlicensed or license-assisted operation of cellular networks is under consideration. For these reasons, we will henceforth refer to the whole $6-24$ \SI{}{\giga\hertz} band as the ``upper midband".}


The high interest in the upper midband necessitates detailed studies of the propagation channels in this frequency range. Determination of the pathloss and shadowing is a basic requirement for assessment of coverage and reliability of a system. Furthermore, the directional dispersion of the channels needs to be investigated; this is all the more important as at these frequency ranges, larger antenna arrays (compared to the sub-6 \SI{}{\giga\hertz} bands) are required to compensate for pathloss. 


Most importantly, these studies need to be performed over the whole bandwidth of interest. Firstly, this serves as basis for spectrum regulators to determine which parts of the spectrum are especially suitable for assignment to cellular services - both with respect to benign propagation conditions for good coverage, and to assess potential coexistence with other services in this band. Secondly, understanding not only the channel in each potential sub-band, but also the correlations between them, forms the basis for band aggregation algorithms. Thirdly, actual \ac{UWB} transmissions that can be used, e.g., for sensing (either over the whole contiguous band, or in non-contiguous sub-bands) can be evaluated from such measurements. And finally, such measurements allow to reliably establish channel models that cover large bandwidths - note that the 3GPP $38.901$ model \cite{3gpp.38.901}, which claims validity between $0.6-100$ \SI{}{\giga\hertz}, is based only on measurements at below 6 \SI{}{\giga\hertz}, and measurements above 24 \SI{}{\giga\hertz}.\footnote{Current activities in 3GPP aim to explicitly consider measurements in FR 3.} Such measurements are inherently \ac{UWB}, which we define here as $>20$ \% relative bandwidth.\footnote{Note that the FCC defines \ac{UWB} as {\em either} $>20$ \% relative bandwidth {\em or} $>500$ MHz absolute bandwidth. However, that latter definition is not necessarily tied to variations of the channel statistics over the measured bandwidth, and is thus of less interest from a channel modeling point of view \cite{molisch2009ultra}.}

\subsection{ Previous results} 

While measurements in the upper midband are much rarer than in either sub-6 \SI{}{\giga\hertz} or millimeter-wave, there still is a considerable number of investigations. Some of those are dedicated to measuring the details of specific propagation effects, such as the transmission through vegetation, e.g., \cite{schwering1988millimeter,abbasi2024vegetationimpact}, transmission through, and reflection from, building materials \cite{shakya2024wideband}, impact of the presence of humans \cite{janssen1996wideband} and cars \cite{rustako1991attenuation}, or diffraction \cite{tervo2014diffraction,ramos2017multiple}. In terms of end-to-end channel measurements, many investigations in this area use SISO (single-input single-output) setups, i.e., a single antenna (which might be omni or directional) at both link ends. These measurements might then be narrowband, e.g., \cite{al2016path,rezende201718,oh2019empirical,yoza2019path}, wideband over one sub-band \cite{salous2014radio,zhou2015experimental}, measurement of one or more bands in the upper midband and comparison with others in lower and higher frequency ranges \cite{medbo2017wireless,diakhate2017millimeter,huang2016propagation} \footnote{We  categorize measurements that are measuring two or more widely separated sub-bands, where each of them is wideband, as wideband, not as \ac{UWB}.} or even \ac{UWB} \cite{kristem2018outdoor} but in any case do not provide the directional information that is required for systems that will use multiple antenna elements. There are also a number measurements that are directionally resolved at one link end, and that are  wideband \cite{fan2016measured,naderpour2016spatio,ling2016comparison} and even \ac{UWB}, though the latter are either indoor, e.g., \cite{muller2016simultaneous,huang2017comparison,wang2022propagation} or rooftop-to-rooftop,  \cite{solomitckii2017comparative}. 

Closest to our work are measurements that are double-directionally resolved. Measurements were done in indoor environments \cite{hui2016channel,dias2016spatial,shakya2024angular,ying2024upper}, and short-distance measurements in an enclosed courtyard \cite{roivainen2016validation}. Measurements in urban microcells include a series of papers on measurements with a switched MIMO channel sounder \cite{saito201711,kim2018multi,saito2017dense} operating at $11$ \SI{}{\giga\hertz}. Other groups measure data throughput and channel characteristics with a $4 \times 4$ systems at $15$ \SI{}{\giga\hertz} \cite{tateishi20165g,okvist201515}, while \cite{suyama2014evaluation} performed transmission experiments with up to $24 \times 24$ MIMO at 11 \SI{}{\giga\hertz}, achieving $30$ Gbit/s throughput.  Ref. \cite{abbasi2025ultrawideband} measures the effect of azimuth angle and elevation diversity in \ac{LoS} channels at $12$ \SI{}{\giga\hertz}. 
Another recent measurement campaign covers both $7$ and $15$ \SI{}{\giga\hertz} \cite{shakya2024urban}, measuring with $1$ \SI{}{\giga\hertz} bandwidth in a microcellular setup. However, none of these outdoor measurements is \ac{UWB}. 
 


There is also a large number of measurements in the FCC-designated \ac{UWB} frequency range $3.1-10.6$ \SI{}{\giga\hertz}, see \cite{molisch2009ultra} and references therein. Most of these measurements are indoors, because \ac{UWB} operation in the band is only allowed for indoor environments, and furthermore SISO. MIMO \ac{UWB} measurements up to $2011$ are surveyed in \cite{molisch2010mimo}, while later such measurements include, e.g., \cite{tsao2012measurements,sangodoyin2015ultrawideband,zahedi2017characterization}. All of these are again in indoor environments, with the exception of  \cite{zahedi2014measurement,al2019path}, which however only measure $3.1-5.3$ \SI{}{\giga\hertz}. Furthermore, all these measurements cover only a small part of the upper midband. The only \ac{UWB} measurement covering large parts of the upper midband, specifically, $3-18$ \SI{}{\giga\hertz}, is the above-mentioned \cite{kristem2018outdoor}, which however is only SISO.


\subsection{Contribution}

To remedy the lack of outdoor double-directional \ac{UWB} measurement campaigns, this paper presents sample results and statistical models for key channel parameters from a measurement campaign in the $6-14$ \SI{}{\giga\hertz} range in an urban microcellular scenario. This is, to the best of our knowledge, the first time such a wideband double-directional analysis has been conducted outdoor over this frequency range. Specific contributions in this paper include:
\begin{itemize}
    \item We present measurement setup and details of the measurement campaign, which analyzes over $25,000$ directional \ac{PDP} to reveal key channel characteristics, where distances between \ac{Tx} and \ac{Rx} range between $60$ and \SI{400}{\meter}.
    \item We analyze sample results both to verify the correctness of the measurements and point out key propagation effects; we particularly note the importance of vegetation attenuation on the link characteristics, and argue that \ac{OLoS} must be treated as separate category from \ac{LoS} and \ac{NLoS} in this frequency range.
    \item We provide statistical characterization of pathloss, delay spread, and angular spread both with respect to distance and as a function of frequency.
\end{itemize}

\subsection{Organization of paper}

The remainder of this paper is organized as follows: Section II describes the measurement setup and environment, followed in Sec. III by a description of the data evaluation procedure and definitions of channel parameters of interest. The main results, both in terms of sample results and statistical evaluations, are presented in Sec. IV. A summary and conclusions wrap up this paper.

\section{Measurement equipment and site}
\subsection{Testbed description}
\begin{figure}[t!]
	\centering
	\includegraphics[width=8.5cm]{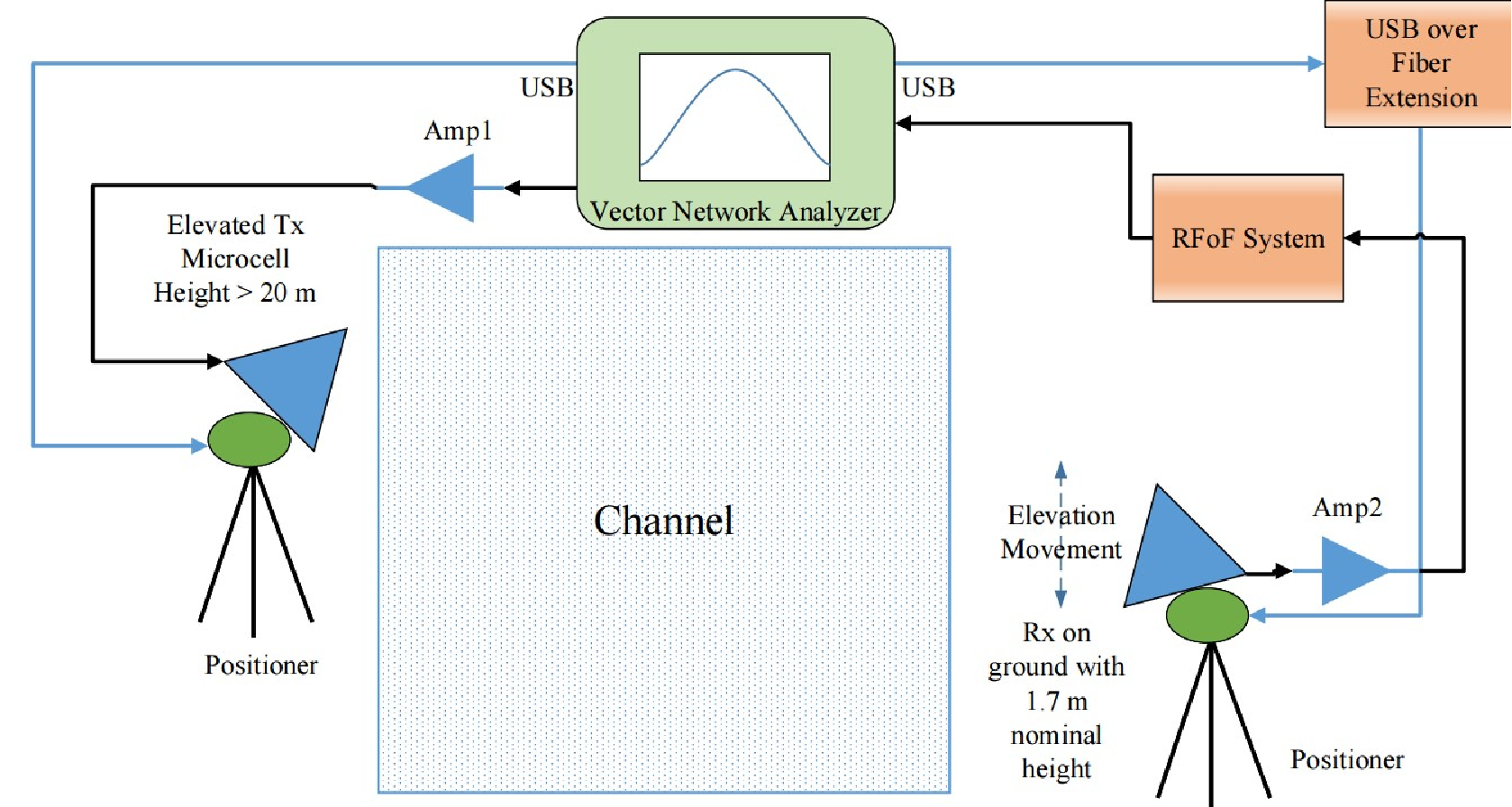}
	\caption{RFoF-based Midband channel measurement setup.}
	\label{fig:setup}
\end{figure}

Measurements with wide bandwidth and directional resolution can be achieved through different combinations of various wideband measurement techniques in the time- or frequency domain, with various types of array measurements \cite[Chapter 9]{molisch2023wireless}. Given the various requirements on bandwidth and directional resolution, combined with constraints on complexity and development time of the setup, we chose a setup that integrates a Vector Network Analyzer (VNA) for covering the operating from \SI{6}{\giga\hertz} to \SI{14}{\giga\hertz} with a pair of HGHA618 horn antennas at the \ac{Tx} (which is placed at the \ac{BS} location and the \ac{Rx}, which is placed at the \ac{UE} location; \ac{BS} and \ac{Tx} are henceforth used interchangeable in this paper for convenience though the channel is reciprocal). Both VNA and horn antennas are \ac{UWB} and thus well suited for the requirements of our measurements. The VNA sequentially measures the channel transfer function at discrete frequency points separated by $1$ MHz, allowing aliasing-free measurement of impulse responses with excess delays up to \SI{1}{\micro\second}. This corresponded to a multipath range of about \SI{300}{\meter}, which was sufficient for the considered environment. Since \ac{Tx} and \ac{Rx} part of the VNA are in the same box, we used a custom radio-frequency-over-fiber (RFoF) connection to haul the signal between \ac{Rx} antenna and the VNA, similar to \cite{abbasi2022thz,abbasi2023thz}. 

To obtain double-directional resolution, the horn antennas are rotated by high-precision mechanical positioners at both \ac{Tx} and \ac{Rx}. In the nominal position, the \ac{Tx} and \ac{Rx} are in the same azimuthal plane i.e., the co-elevation is defined such that $0^\circ$ corresponds to the LOS direction (i.e. \ac{Tx} looks down whereas the \ac{Rx} looks up). The \ac{Tx} was held at a single elevation angle, covering azimuth angles from $-60^\circ$ to $60^\circ$ in steps of $10^\circ$. Meanwhile, the \ac{Rx} traversed five co-elevation angles from $-20^\circ$ to $20^\circ$ in increments of $10^\circ$ and swept from $0^\circ$ to $360^\circ$ in $10^\circ$ steps. As a result, each \ac{Tx}-\ac{Rx} location pairing yielded 2340 measured transfer functions. Due to the duration of the mechanical movements and required stabilization/settling periods after each rotation step, a full azimuth-elevation scan required several hours. Consequently, measurements were primarily conducted at night to limit environmental motion and disturbances. Pedestrian access was also restricted to maintain near-static conditions, though wind-induced motion of foliage could still occur.    

Table \ref{tab:parameters} summarizes the main measurement parameters.

\begin{table}[t!]
	\centering
	\caption{Setup parameters.}
	\label{tab:parameters}
	\begin{tabular}{|l||c|c|}
		\hline
		\textbf{Parameter}              & \textbf{Symbol}   & \textbf{Value} \\ \hline\hline
		\textit{Frequency points}     & $N$                 & 8001           \\
		\textit{Tx height}     & $h_{Tx}$                 & 20.38 m           \\
		\textit{Rx height}     & $h_{Rx}$                 & 1.7 m           \\
		\textit{Start frequency}        & $ f_{start} $     & \SI{6}{\giga\hertz}          \\
		\textit{Stop frequency}         & $ f_{stop} $      & \SI{14}{\giga\hertz}         \\
		\textit{Bandwidth}              & $BW$              & \SI{8}{\giga\hertz}        \\	
		\textit{IF bandwidth}              & $IF_{BW}$              & 1 KHz         \\
		\textit{Tx Az rotation range}   & $\phi_{\rm Tx}$        & [$-$60$^{\circ}$,60$^{\circ}$]           \\ 
		\textit{Tx Az rotation resolution}   & $\Delta \phi_{\rm Tx}$        & 10$^{\circ}$           \\
		\textit{Rx Az rotation range}   & $\phi_{\rm Rx}$        & [0$^{\circ}$,360$^{\circ}$]           \\
		\textit{Rx Az rotation resolution}   & $\Delta \phi_{\rm Rx}$        & 10$^{\circ}$           \\
		\textit{Rx El rotation range}   & $\tilde{\theta}_{\rm Rx}$        & [$-$20$^{\circ}$,20$^{\circ}$]           \\
		\textit{Rx El rotation resolution}   & $\Delta \tilde{\theta}_{\rm Rx}$        & 10$^{\circ}$           \\ \hline
	\end{tabular}
\end{table}

To isolate the inherent system and antenna response (gain removed through directional filtering effects remain) from the channel, a time-gated \ac{OTA} calibration was performed daily at a reference point with clear \ac{LoS}, approximately \SI{56.45}{\meter} from the \ac{Tx}.  

\subsection{Measurement scenario}
\begin{figure}[t!]
	\centering
	\includegraphics[width=8.5cm]{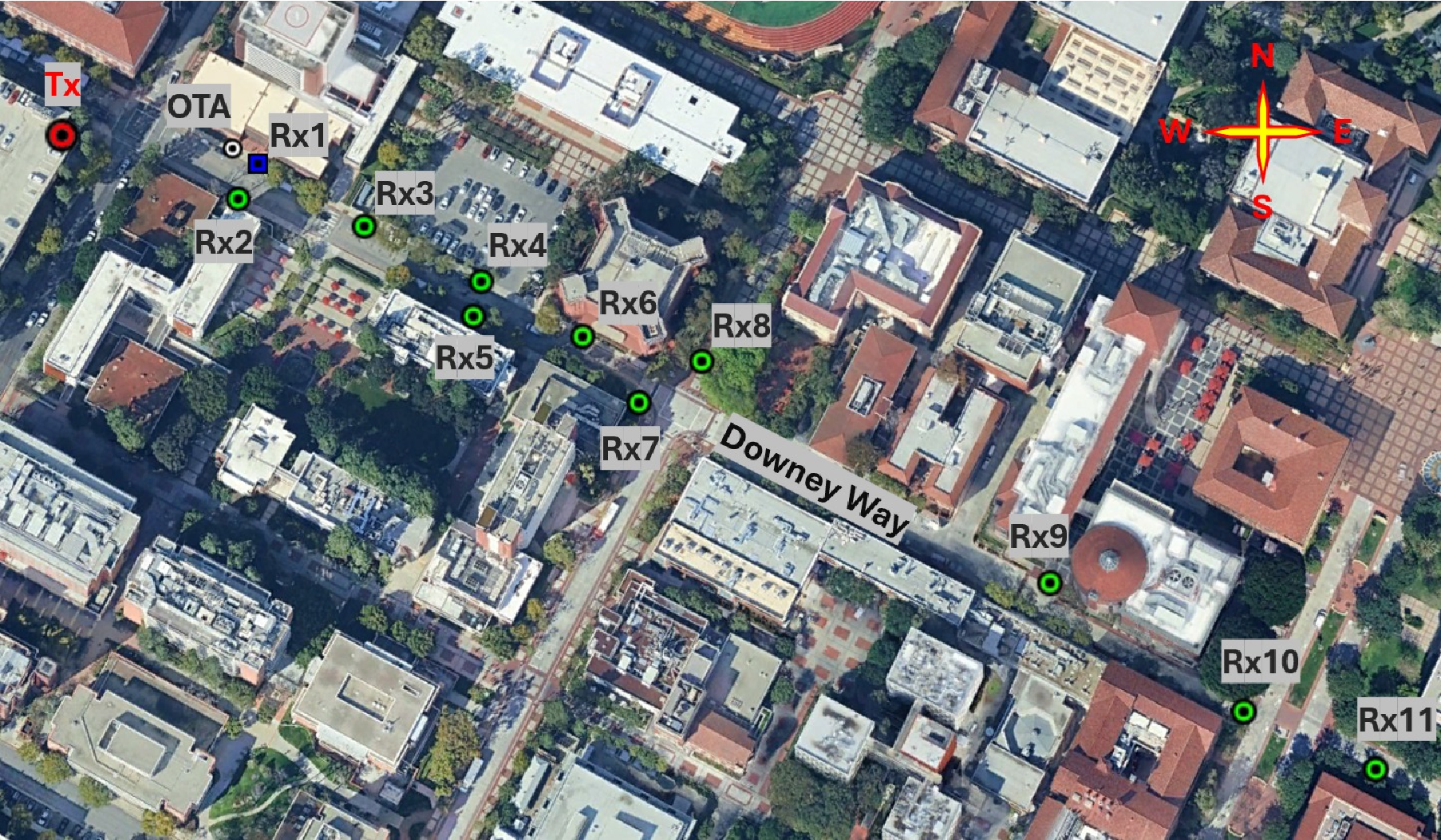}
	\caption{Measurement scenario.}
	\label{fig:scenario}
\end{figure}

The channel measurement study was carried out in an urban microcell environment, featuring a mix of buildings, streets, and open spaces, specifically the University Park Campus (UPC) of the University of Southern California in Los Angeles, California, USA. The \ac{Tx} was placed at a height of \SI{20.38}{\meter} above ground level on an external staircase of the Downey Way Parking Structure (PSA). The 11 \ac{Rx} points were at a height of \SI{1.7}{\meter} and placed along Downey Way (the street on which PSA is located), such that no buildings obstructed the \ac{LoS} connection between \ac{Tx} and \ac{Rx}. The Rx1 is located at a distance of \SI{65.1}{\meter} from the \ac{Tx}, ensuring clear \ac{LoS}, while the longest link, Rx11, was positioned \SI{436.1}{\meter} away, marking the farthest channel in the measurement campaign (see Table \ref{tab:distances} for all distances). Vegetation of varying density was present across the site, representing a significant source of signal blockage in certain regions; points Rx2 - Rx11 were all classified as suffering from \ac{OLoS} propagation conditions. Apart for the $0^\circ$ co-elevation described earlier, the nominal azimuthal position of the \ac{Tx} and all the \ac{Rx} is parallel to Downey Way (\ac{Tx} turned South-East and \ac{Rx}s turned towards North-West).

\begin{table}[t!]
\centering
\caption{Distances corresponding to each Rx Identifier.}
\label{tab:distances}
\begin{tabular}{|c||c|}
\hline
\textbf{Rx Identifier} & \textbf{Distance (m)} \\ \hline \hline
Rx1 & 65.1 \\
Rx2 & 62.1 \\
Rx3 & 103.5 \\
Rx4 & 139.1 \\
Rx5 & 143.6 \\
Rx6 & 162.8 \\
Rx7 & 201.4 \\
Rx8 & 214.9 \\
Rx9 & 336.3 \\
Rx10 & 404.9 \\
Rx11 & 436.1 \\ \hline
\end{tabular}
\end{table}

\section{Parameters and processing}
\subsection{Data Processing}
The VNA-based measurement setup produces frequency scans at various \ac{Tx}-\ac{Rx} locations, forming a four-dimensional tensor $H_{meas}(f,\phi_{\rm Tx},\phi_{\rm Rx},\tilde{\theta}_{\rm Rx};d)$, where $f$ is the frequency, $\phi_{\rm Tx}$ and $\phi_{\rm Rx}$ represent \ac{Tx} and \ac{Rx} azimuth orientations, $\tilde{\theta}_{\rm Rx}$ denotes the \ac{Rx} co-elevation, and $d$ is the \ac{Tx}-\ac{Rx} distance. The calibrated channel transfer function is obtained by dividing $H_{meas}$ by the \ac{OTA} calibration, $H_{OTA}(f)$:   
\begin{align}
    H(f,\phi_{\rm Tx},\phi_{\rm Rx},\tilde{\theta}_{\rm Rx};d) &= 
    \frac{H_{meas}(f,\phi_{\rm Tx},\phi_{\rm Rx},\tilde{\theta}_{\rm Rx};d)}
    {H_{OTA}(f)}.
\end{align}
The \ac{OTA} calibration removes the effect of the measurement system, including antenna gain (while retaining directional filtering effects), and the impact of measuring at the \ac{OTA} distance, $d_{OTA} = \SI{56.45}{\meter}$.   

The directional \ac{PDP}s are computed as:
\begin{multline}
    P_{calc}(\tau,\phi_{\rm Tx},\phi_{\rm Rx},\tilde{\theta}_{\rm Rx},d) = \\ 
    \left| \mathcal{F}_{f}^{-1} \left\{ H(f,\phi_{\rm Tx},\phi_{\rm Rx},\tilde{\theta}_{\rm Rx},d) \right\} 
    \right|^2.      
\end{multline}  

Noise reduction is achieved using thresholding and delay gating, similar to \cite{gomez-ponce2020,abbasi2021double}, and is defined as:   
\begin{align}  
    P(\tau) &= \left[ P_{calc}(\tau) : (\tau \leq \tau_{gate}) \land 
    (P_{calc}(\tau) \geq P_{\lambda}) \right],
\end{align}
where $\tau_{gate}$ is set to 966.67 ns (corresponding to 290 m excess runlength) to avoid using long delay bins or points with the ``wrap-around" effect of the IFFT. Meanwhile, $P_{\lambda}$ is fixed at \SI{22}{\decibel} below the maximum power received in a \ac{PDP}. The \SI{22}{\decibel} cutoff is chosen as it represents the lowest dynamic range observed across any sub-band during the measurements, and by choosing the same dynamic range across all points, we want to ensure consistent and fair evaluation. For further discussion on the impact of selecting noise and delay thresholds, refer to \cite{gomez2023impact}.   

The strongest directional \ac{PDP} (Max-Dir) is selected as the beam-pair with the highest power:  
\begin{align}
    P_{\rm Max-Dir}(\tau)
     &= \arg\max_{i,j,k} \sum_\tau 
    P(\tau,\phi_i,\phi_j,\tilde{\theta}_k,d).
\end{align}

\begin{figure}[t!]
	\centering
	\includegraphics[width=0.9\columnwidth]{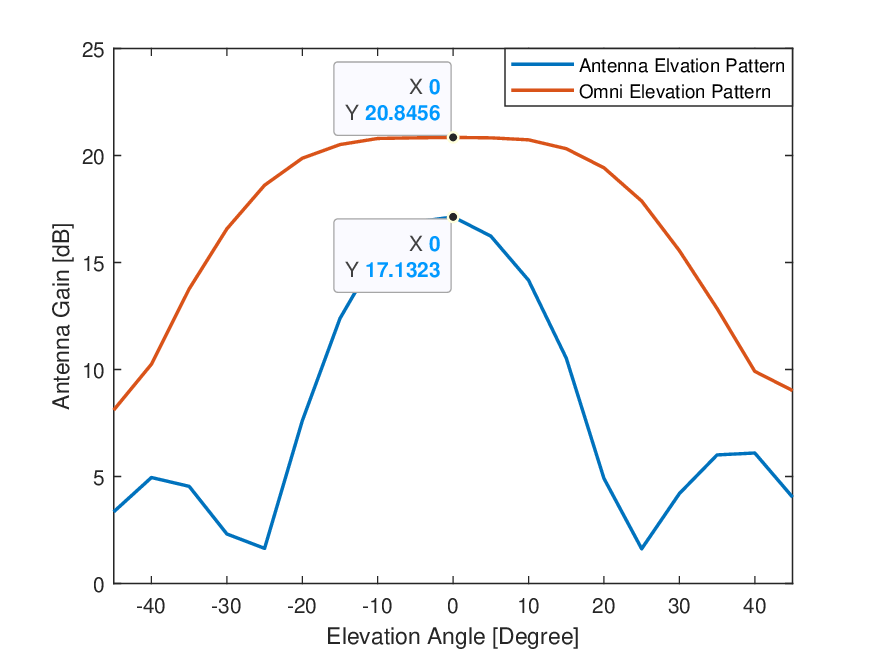}
	\caption{Elevation pattern for the original horn and modified omni at \SI{6.5}{\giga\hertz}.}
	\label{fig:antenna}
\end{figure}

An omni-directional \ac{PDP} is constructed by first adding up the directional PDPs for different co-elevations, and then selecting the strongest azimuth direction per delay bin, similar to \cite{Hur_omni}:   
\begin{align}
    P_{\rm omni}'(\tau;d) &= \max_{\phi_{\rm Tx},\phi_{\rm Rx}} \sum_k 
    P(\tau,\phi_{\rm Tx},\phi_{\rm Rx},\tilde{\theta}_{k};d),
\end{align}
where $k \in \{1, 2, 3, 4, 5\}$ represents \ac{Rx} co-elevations ($-20^\circ$ to $20^\circ$), spaced in $10^\circ$ increments. Summing the PDP over the different elevations can be interpreted as measuring with a modified antenna pattern that is broader, and also has a higher peak gain, than the original horn pattern. The difference in gain (on a \SI{}{\decibel} scale) between this virtual pattern and the original horn pattern, which depends on frequency, thus is subtracted from all $P_{\rm omni}'(\tau;d)$ to obtain the $P_{\rm omni}(\tau;d)$ that are then used for further processing.  To show this, we have plotted the elevation pattern for the HGHA618 antenna and the respective omni construction at \SI{6.5}{\giga\hertz} in Fig. \ref{fig:antenna}. We see that at an additional gain of \SI{3.7}{\decibel} in the omni pattern. This value reduces for the current antenna when we go higher in frequency and at \SI{13.5}{\giga\hertz} it is \SI{1.57}{\decibel}.

\subsection{Parameter Computation}
With the calibrated impulse responses, we derive several key channel parameters, as described below.   

\subsubsection{Path Loss and Shadowing}
Path gain is defined as the sum of power across all delay bins in the \ac{PDP}:  
\begin{align}
    PG_i(d) &= \sum_\tau P_i(\tau,d),
\end{align}
where $i$ represents either omni-directional or strongest beam. Path loss is calculated as $PL_i(d) = PG^{-1}_i(d)$. On a logarithmic scale, path loss is fitted to the power law model:  
\begin{align}
    PL_{\rm dB}(d) &= \alpha + 10\beta \log_{10}(d) + \epsilon,
\end{align}   
where $\alpha$ and $\beta$ are the estimated parameters, and $\epsilon$ represents shadowing, modeled as $\epsilon \sim N(0,\sigma)$ \cite{molisch2023wireless, kartunen_PL}. Weighting techniques from \cite{karttunen2016path} are applied to address non-uniform distance density. 

\subsubsection{Delay Spread}
The \ac{RMSDS} is the square root of the second central moment of the \ac{PDP} \cite{molisch2023wireless}:  
\begin{align}
    \sigma_\tau &= \sqrt{\frac{\int_\tau P_i(\tau) \tau^2 \, d\tau}{\int_\tau P_i(\tau) \, d\tau} - 
    \left( \frac{\int_\tau P_i(\tau) \tau \, d\tau}{\int_\tau P_i(\tau) \, d\tau} \right)^2},
\end{align}  
where $i$ denotes either ``omni" or ``Max-Dir". A Hann window is applied to $H(\cdot)$ in (2) \cite{payami2013delay} to reduce the sidelobes that might affect the \ac{RMSDS}. Also, as \ac{RMSDS} applies to continuous waveforms, we approximate it by oversampling by a factor of 10. Finally, for consistency with channel modeling literature, the \ac{RMSDS} is expressed on a logarithmic scale in \SI{}{\decibel\second} as $10 \log_{10} (\text{Delay Spread}/\SI{1}{\second})$.  

\subsubsection{Angular Spread}  
The \ac{AS} is derived from the double-directional angular power spectrum ($DDAPS_{\rm full}$), calculated as \cite[Chapter 6]{molisch2023wireless}:   
\begin{multline}
    DDAPS_{\rm full}(\phi_{\rm Tx},\phi_{\rm Rx},\tilde{\theta}_{\rm Rx};d) = \\ 
    \sum_\tau P(\tau,\phi_{\rm Tx},\phi_{\rm Rx},\tilde{\theta}_{\rm Rx};d),
\end{multline}
which, after summing over co-elevations, becomes:
\begin{multline}
    DDAPS(\phi_{\rm Tx},\phi_{\rm Rx};d) = \\\sum_{\tilde{\theta}_{\rm Rx}} 
    DDAPS_{\rm full}(\phi_{\rm Tx},\phi_{\rm Rx},\tilde{\theta}_{\rm Rx};d).
\end{multline}   
Fleury's definition \cite{fleury2000first} is used to compute the angular spread:
\begin{align}
    \sigma^\circ &= \sqrt{\frac{\sum_\phi \left| e^{j\phi} - \mu_\phi \right|^2 APS_o(\phi)}
    {\sum_\phi APS_o(\phi)}},
\end{align}  
where $o$ denotes \ac{Tx} or \ac{Rx}, and $\mu_\phi$ is the mean angular value:     
\begin{align}
    \mu_\phi &= \frac{\sum_\phi e^{j\phi} APS_o(\phi)}{\sum_\phi APS_o(\phi)}.  
\end{align} 
Before we proceed, we highlight that the definition of \ac{AS} used here results in values ranging between 0 and 1. For small $\sigma^\circ$ values (up to $0.3$), this corresponds approximately to the angular spread as computed via the second central moment in radians. Additionally, we note that the lower bound of the measured \ac{AS} is determined by the pattern of the directional antenna, as discussed in \cite{Abbasi2021THz}. 



Both \ac{RMSDS} and \ac{AS} are modeled using two approaches: (i) modeling of the cumulative distribution functions (cdfs) at all measurement locations; these are modeled as normal distribution (on a dB scale), characterized by their mean ($\mu$) and standard deviation ($\sigma$); (ii) modeling of their distance dependence, as a linear regression modeling in the form $\alpha + 10\beta \log_{10}(d)$. Confidence intervals are provided for both approaches to offer a comprehensive understanding of the observed trends and variability.

\section{Measurement results}
In this section, we discuss the results of our measurement campaign, focusing on key observations and analyses derived from the data. First, the \ac{PDP}s for selected \ac{LoS} and \ac{OLoS} cases are examined to highlight the delay dispersion characteristics of the measured channels. Next, the \acp{APS} are analyzed to understand the spatial distribution of energy, including directional characteristics and reflections. We then present statistical modeling results, in particular \ac{PL} modeling results, comparing omni-directional and Max-Dir cases, along with shadowing statistics. Subsequently, the \ac{RMSDS} is evaluated to characterize delay dispersion as a function of distance. Finally, \ac{AS} measurements are explored, detailing their dependence on \ac{Tx} and \ac{Rx} orientations and distance. In all these investigations, we pay particular attention to the frequency dependence of the analyzed parameters.
\begin{figure*}[t!]
\centering
    \begin{subfigure}{0.4\textwidth}
            \centering
            \hspace{7mm}
            \includegraphics[width=1\columnwidth]{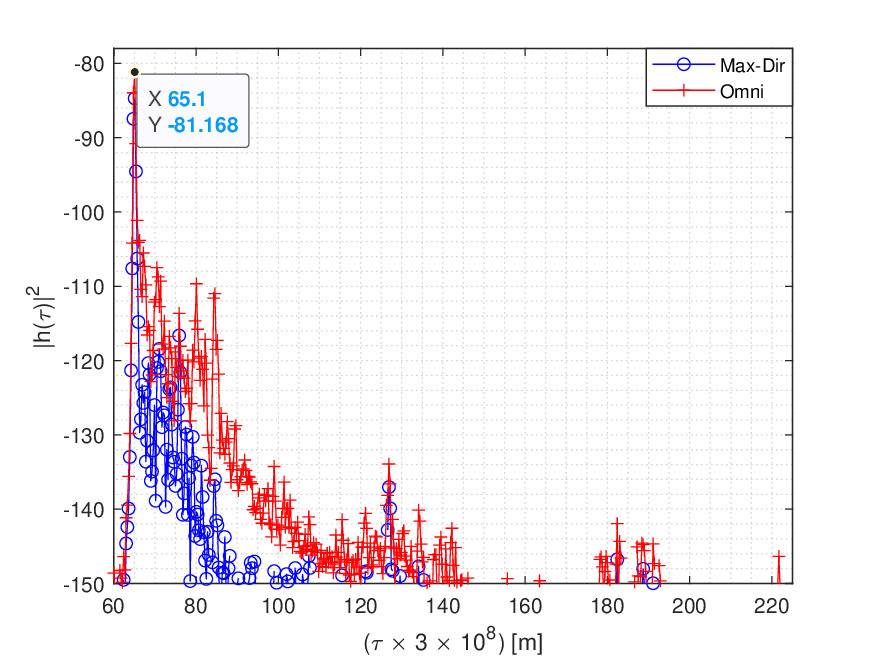}
            \caption{$6-7$ \SI{}{\giga\hertz} LoS case (\ac{Rx}1) with $d=65.1 m$.}
            \label{fig:PDP1}
    \end{subfigure}
    \begin{subfigure}{0.4\textwidth}
    \centering
        \centering
        \hspace{7mm}
        \includegraphics[width=1\columnwidth]{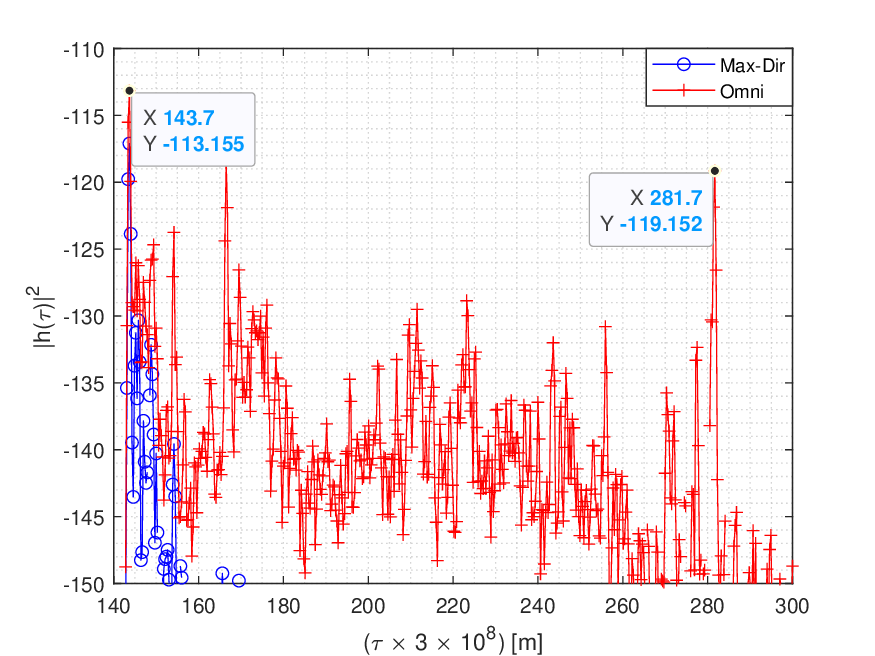}
        \caption{$6-7$ \SI{}{\giga\hertz} OLoS case (\ac{Rx}5) with $d=143.6 m$.}
        \vspace*{0mm}
        \label{fig:PDP2}
    \end{subfigure}
    \begin{subfigure}{0.4\textwidth}
            \centering
            \hspace{7mm}
            \includegraphics[width=1\columnwidth]{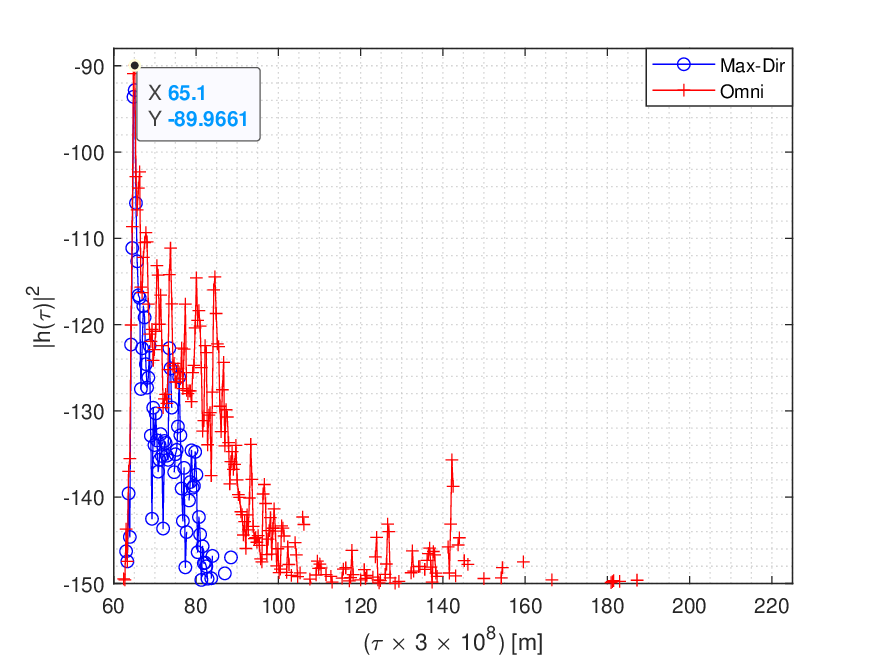}
            \caption{$13-14$ \SI{}{\giga\hertz} LoS case (\ac{Rx}1) with $d=65.1 m$.}
            \label{fig:PDP3}
    \end{subfigure}
    \begin{subfigure}{0.4\textwidth}
    \centering
        \centering
        \hspace{7mm}
        \includegraphics[width=1\columnwidth]{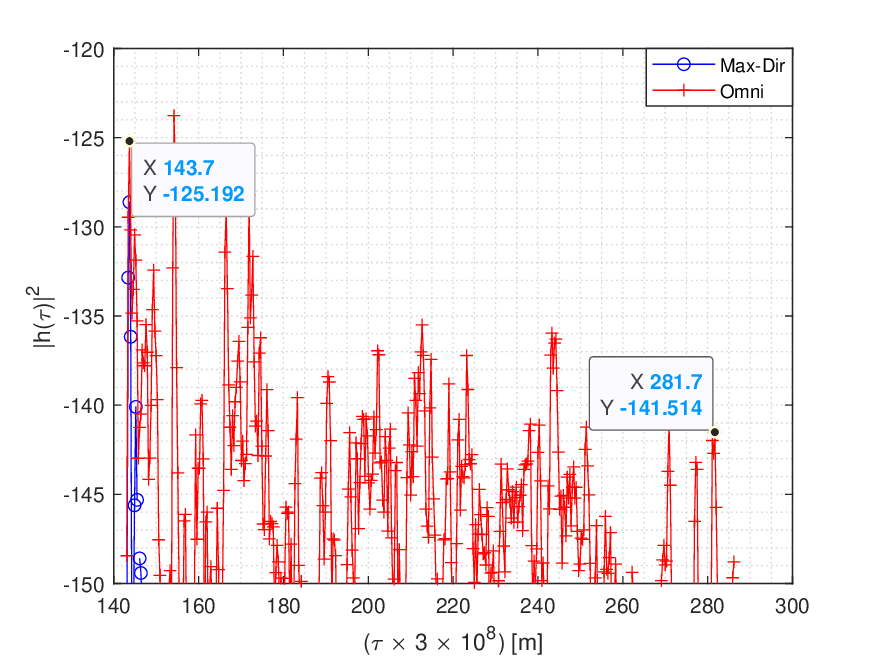}
        \caption{$13-14$ \SI{}{\giga\hertz} OLoS case (\ac{Rx}5) with $d=143.6 m$.}
        \vspace*{0mm}
        \label{fig:PDP4}
    \end{subfigure}
    \caption{\ac{PDP} for two sample measurement cases. The symbols (crosses, circles) only serve to distinguish the Max-Dir and omni cases, and do not signify multipath components.}
    \label{fig:PDP}
\end{figure*}
\subsection{Power delay profiles}
In this subsection, we examine the \ac{PDP}s for two representative cases: \ac{LoS} (Rx1) at $d=\SI{65.1}{\meter}$, and \ac{OLoS} (Rx5) at $d=\SI{143.6}{\meter}$, for two frequency bands - $6-7$ \SI{}{\giga\hertz} and $13-14$ \SI{}{\giga\hertz} - as illustrated in Fig. \ref{fig:PDP}. Note that the delay on the x-axis in these figures is given in meters, corresponding to the runlength (calculated as $\tau \times c$, where $c$ is the speed of light, $\SI{3e8}{\meter\per\second}$). This transformation facilitates a more intuitive analysis by directly mapping the delay to spatial distances. 

We note that in all cases, the first \ac{MPC} corresponds to the \ac{LoS} or primary propagation component, as expected, appearing at a delay corresponding to the \ac{Tx}-\ac{Rx} distance. 
For Rx1, the \ac{LoS} component is clearly visible at $d=\SI{65.1}{\meter}$ for both bands (Fig. \ref{fig:PDP1} and Fig. \ref{fig:PDP3}). Several additional reflections from nearby structures contribute to the delay spread, with many arriving shortly after the \ac{LoS} component due to their relatively short additional travel distances. 
The omni-directional \ac{PDP} for Rx1 captures these reflections from multiple directions, including strong reflections from buildings across Downey Way, resulting in a higher number of \ac{MPC}s compared to the Max-Dir \ac{PDP}. The Max-Dir \ac{PDP} is more concentrated near the \ac{LoS} component, as it captures fewer \ac{MPC}s due to the spatial filtering provided by directional antennas, which restrict the range of \ac{DoA}s and \ac{DoD}s. The \ac{LoS} component is more than $\SI{25}{\decibel}$ stronger than any other component in the omnidirectional PDP, and about $\SI{35}{\decibel}$ stronger than any other component in the Max-Dir PDP for the $6-7$ \SI{}{\giga\hertz} band in Fig. \ref{fig:PDP1}. For the $13-14$ \SI{}{\giga\hertz} band in Fig. \ref{fig:PDP3}, it is nearly $\SI{20}{\decibel}$ stronger than any other component in the omnidirectional PDP, and about $\SI{30}{\decibel}$ stronger than any other component in the Max-Dir PDP.

\begin{figure}[t!]
	\centering\includegraphics[width=8.5cm]{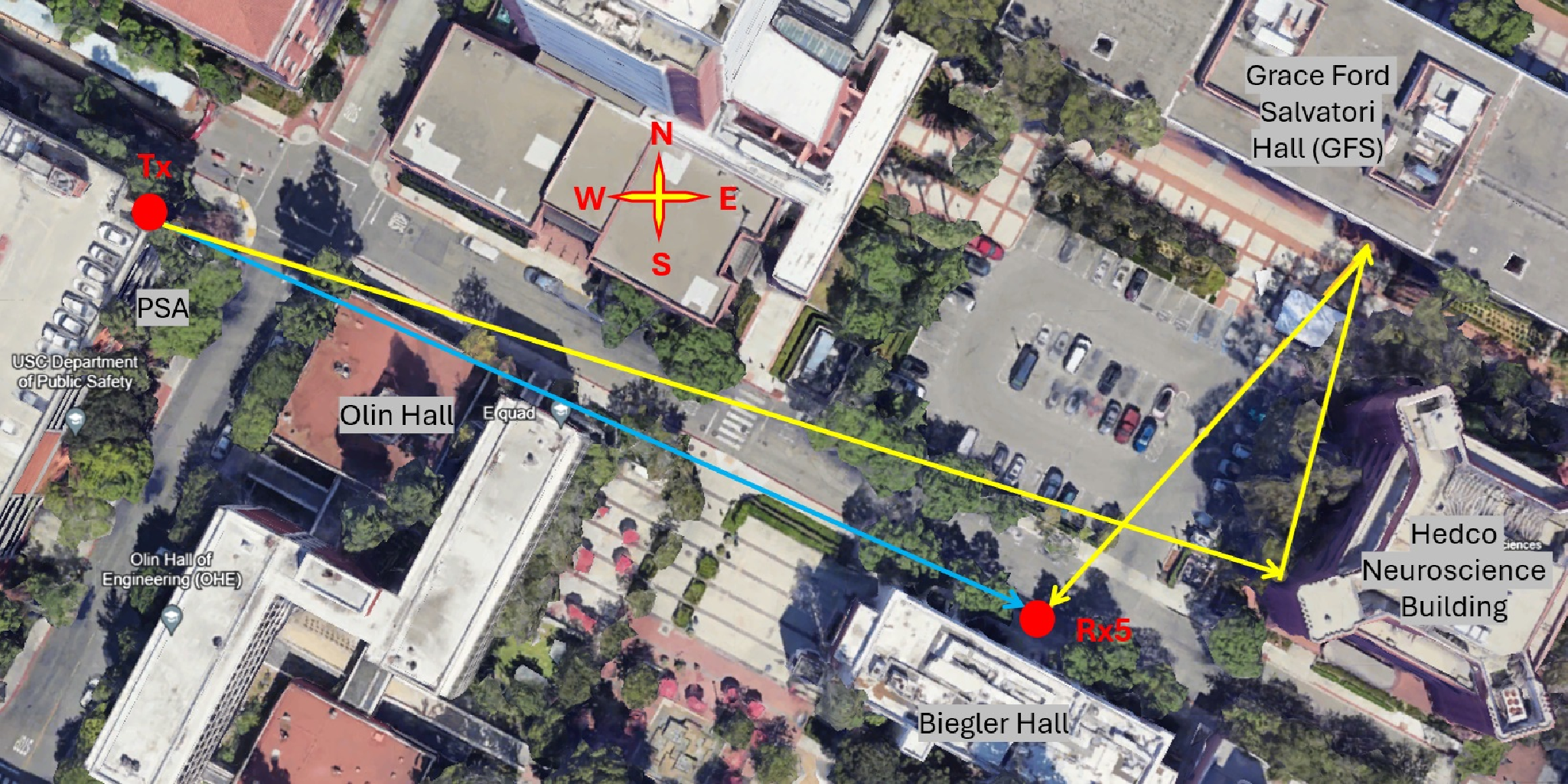}
	\caption{Measurement scenario for Rx5.}
	\label{fig:scenario_rx5}
\end{figure}

The received power for Rx1 derived from the \ac{PDP} is $\SI{-84.88}{\decibel}$ in the $6-7~\SI{}{\giga\hertz}$ band, and $\SI{-91.54}{\decibel}$ in the $13-14~\SI{}{\giga\hertz}$ band after removing the additional omni-directional gain of \SI{3.7}{\decibel} at \SI{6.5}{\giga\hertz}, and \SI{1.57} {\decibel} at \SI{13.5}{\giga\hertz}, respectively ($-81.17~-~3.71~=~-84.88~\SI{}{\decibel}$ ; $-89.97~-~1.57~=~-91.54~\SI{}{\decibel}$).
These values are quite near those generated by Friis free-space model ($\SI{-84.97}{\decibel}$ at $\SI{6.5}{\giga\hertz}$, and $\SI{-91.32}{\decibel}$ at $\SI{13.5}{\giga\hertz}$). The small differences arise from the discretization of the delay axis where sometimes more components fall in a delay bin and at other times where some part of a component is between delay bins.
Also, note that some delayed components around $\SI{220}{\meter}$, though quite small in power, completely vanish at the higher frequency alluding to the fact that the dynamic range is changing across frequencies.

For Rx5, an \ac{OLoS} scenario, the first component at $d=\SI{143.7}{\meter}$ also corresponds to the direct propagation path, which is obstructed by vegetation in Fig. \ref{fig:PDP2} and Fig. \ref{fig:PDP4}. This obstruction causes excess attenuation. The received power for Rx5 ($\SI{-116.87}{\decibel}$ at $6-7~\SI{}{\giga\hertz}$, and $\SI{-126.76}{\decibel}$ at $13-14~\SI{}{\giga\hertz}$ after removing additional omni gain as discussed in Sec. III.A) is lower than the Friis-predicted received power of $\SI{-91.85}{\decibel}$ at \SI{6.5}{\giga\hertz}, and $\SI{-98.20}{\decibel}$ at \SI{13.5}{\giga\hertz}, primarily due to the excess loss introduced by the vegetation blocking the main path. Similar to the previous case, we also note here that the dynamic range of the measurement seems to decrease for the higher frequency band, though it must be noted that this does not effect all the \ac{MPC}s in the same way with some components suffering higher losses showing that the channel is changing with the frequency (note that while the statistical parameters are computed with a fixed dynamic range as discussed in Sec. III, Fig. 3 uses a threshold with fixed absolute level). 
\begin{figure*}[t!]
\centering
    \begin{subfigure}{0.4\textwidth}
            \centering
            \hspace{7mm}
            \includegraphics[width=1\columnwidth]{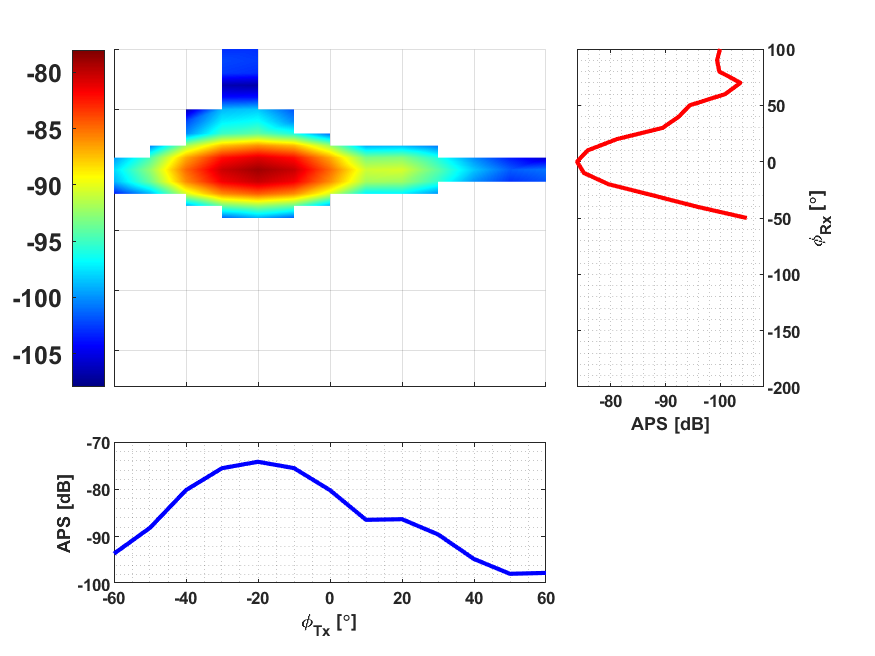}
            \caption{$6-7$ \SI{}{\giga\hertz} LoS case (\ac{Rx}1) with $d=65.1 m$.}
            \label{fig:APS1}
    \end{subfigure}
    \begin{subfigure}{0.4\textwidth}
    \centering
        \centering
        \hspace{7mm}
        \includegraphics[width=1\columnwidth]{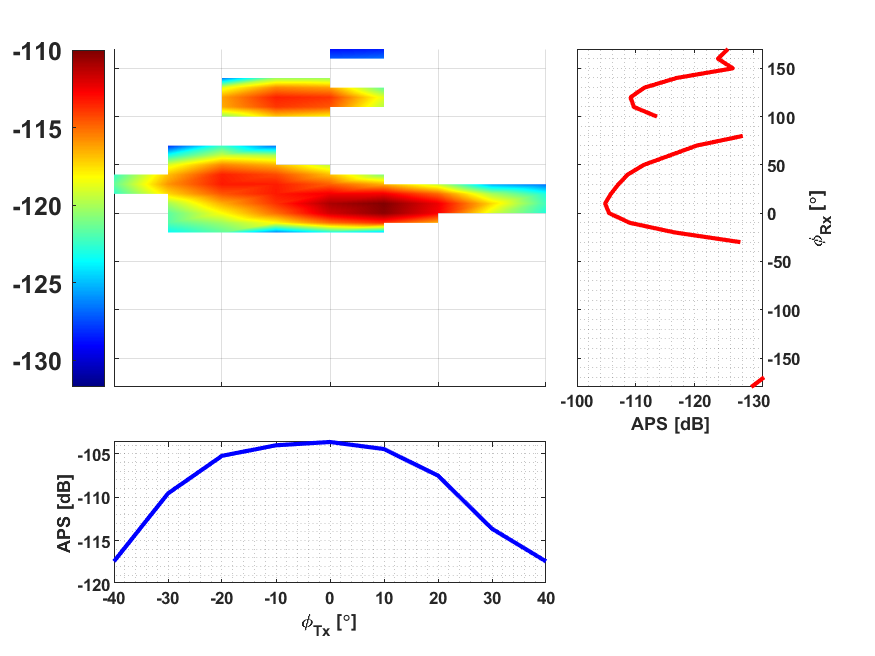}
        \caption{$6-7$ \SI{}{\giga\hertz} OLoS case (\ac{Rx}5) with $d=143.6 m$.}
        \vspace*{0mm}
        \label{fig:APS2}
    \end{subfigure}
    \begin{subfigure}{0.4\textwidth}
            \centering
            \hspace{7mm}
            \includegraphics[width=1\columnwidth]{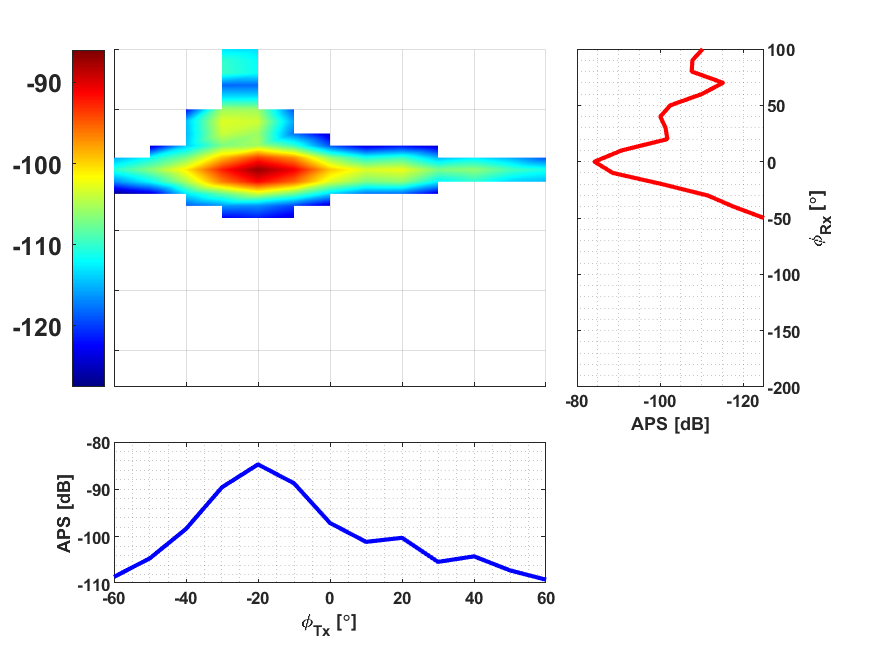}
            \caption{$13-14$ \SI{}{\giga\hertz} LoS case (\ac{Rx}1) with $d=65.1 m$.}
            \label{fig:APS3}
    \end{subfigure}
    \begin{subfigure}{0.4\textwidth}
    \centering
        \centering
        \hspace{7mm}
        \includegraphics[width=1\columnwidth]{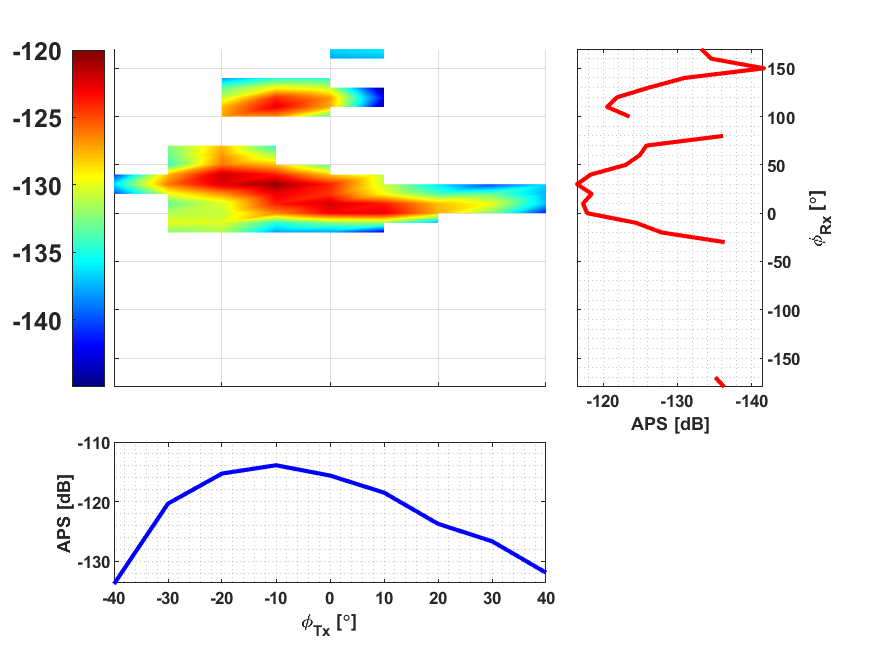}
        \caption{$13-14$ \SI{}{\giga\hertz} OLoS case (\ac{Rx}5) with $d=143.6 m$.}
        \vspace*{0mm}
        \label{fig:APS4}
    \end{subfigure}
    \caption{\ac{APS} for two sample measurement cases.}
    \label{fig:APS}
\end{figure*}

Fig. \ref{fig:scenario_rx5} provides an overview of the measurement scenario for Rx5, showing the location of the \ac{Tx}, Rx5, and key propagation paths. 
We highlight two important paths contributing to the overall \ac{PDP}. The first is the direct path (blue line in Fig. \ref{fig:scenario_rx5}), which is not obstructed by Olin Hall, as the building is lower than the \ac{Tx}. The second is a double-reflected path (yellow line in Fig. \ref{fig:scenario_rx5}) with a total path length of \SI{281}{\meter}. This path involves reflections first from the facade of the Hedco Neuroscience Building and then from the GFS Building before reaching the \ac{Rx}. These multipath contributions, while lower in power compared to the direct path, still play a role in shaping the \ac{PDP}. As a result of the vegetation loss, the main component now has similar power in comparison to other reflected components; for example, the \SI{281}{\meter} delay component is only $\SI{6}{\decibel}$ weaker than the \ac{LoS} component despite suffering reflection losses at two surfaces in the $6-7~\SI{}{\giga\hertz}$. The same path is nearly $\SI{16}{\decibel}$ below the main component for the $13-14~\SI{}{\giga\hertz}$ pointing to the channel's frequency selectivity. 
This is unlike the \ac{LoS} case, where the main component is nearly always much higher in power when compared to the reflections. The attenuation of the direct component due to vegetation is consistent with prior work on vegetation-induced losses in upper-midband frequencies, as detailed in \cite{abbasi2024vegetationimpact}.   

\subsection{Angular power spectrum}
Next we analyze the \ac{APS} for the same two sample points we discussed in detail from a delay perspective in the previous subsection: \ac{LoS} (Rx1) and \ac{OLoS} (Rx5) with results shown in Fig. \ref{fig:APS}. The \ac{APS} provides additional insights into the angular distribution of energy, highlighting both dominant and reflected multipath components. 

Fig. \ref{fig:APS1} and Fig. \ref{fig:APS3}  show the \ac{APS} of LoS point Rx1 at $6-7$ \SI{}{\giga\hertz} and  $13-14$ \SI{}{\giga\hertz}, respectively. The main component is centered at $-20^\circ$ at the \ac{Tx} and $0^\circ$ at the \ac{Rx}, corresponding to the point being located on the northern side of Downey Way. This main \ac{LoS} component is significantly stronger than the other components, as also observed in the \ac{PDP} for Rx1. The reflected components appear at various angles, but their powers are much lower compared to the \ac{LoS} component. These reflections originate from surrounding structures, contributing to the angular dispersion, but their relative power is not substantial enough to influence the overall angular profile significantly. Note, however, that far-away components get weighted quadratically for the \ac{AS}, so that they can have a more pronounced effect there. Across both the bands, we see that the higher frequency band in Fig. \ref{fig:APS3}, has a visibly smaller angular spread, which we would expect based on the results of the \ac{PDP}s.

Fig. \ref{fig:APS2} and Fig. \ref{fig:APS4} show the \ac{APS} of the \ac{OLoS} point Rx5 at $6-7$ \SI{}{\giga\hertz} and  $13-14$ \SI{}{\giga\hertz}, respectively. The main reflections are centered around $0^\circ$ at the \ac{Tx} and $10^\circ$ at the \ac{Rx}, indicating the dominant propagation path. The larger relative power of the reflected \acp{MPC} is visible also in the \ac{APS}. Notably, reflections from the GFS Building, as discussed in the previous subsection, are observed as a distinct cluster centered around $120^\circ$ at the \ac{Rx}. Such additional clusters arising at different angles, impact both diversity and spatial multiplexing, as well as potential inter-user interference in multi-user MIMO settings; due to the near-far effect, interference even from clusters with significantly weaker absolute strength can lead to  inter-user interference whose effect can be significant depending on both its strength and the particular receiver structure \cite{gomez2023impact}.

\subsection{Path loss and shadowing}

\begin{figure*}[t!]
\centering
    \begin{subfigure}{0.4\textwidth}
            \centering
            \hspace{7mm}
            \includegraphics[width=1\columnwidth]{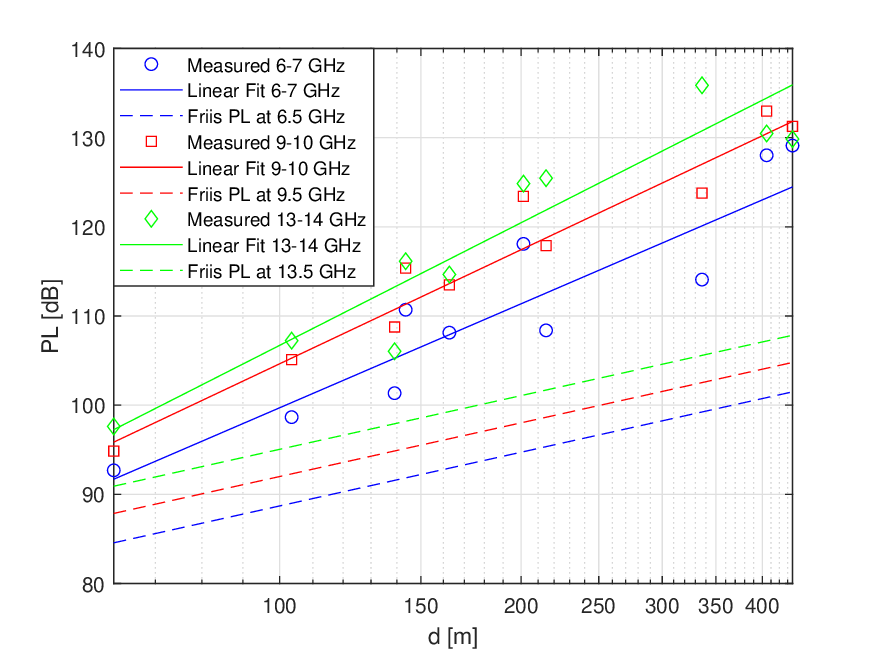}
            \caption{\ac{PL} modeling with $log_{10}(d)$ weighting.}
            \label{fig:PL_Omni1}
    \end{subfigure}
    \begin{subfigure}{0.4\textwidth}
    \centering
        \centering
        \hspace{7mm}
        \includegraphics[width=1\columnwidth]{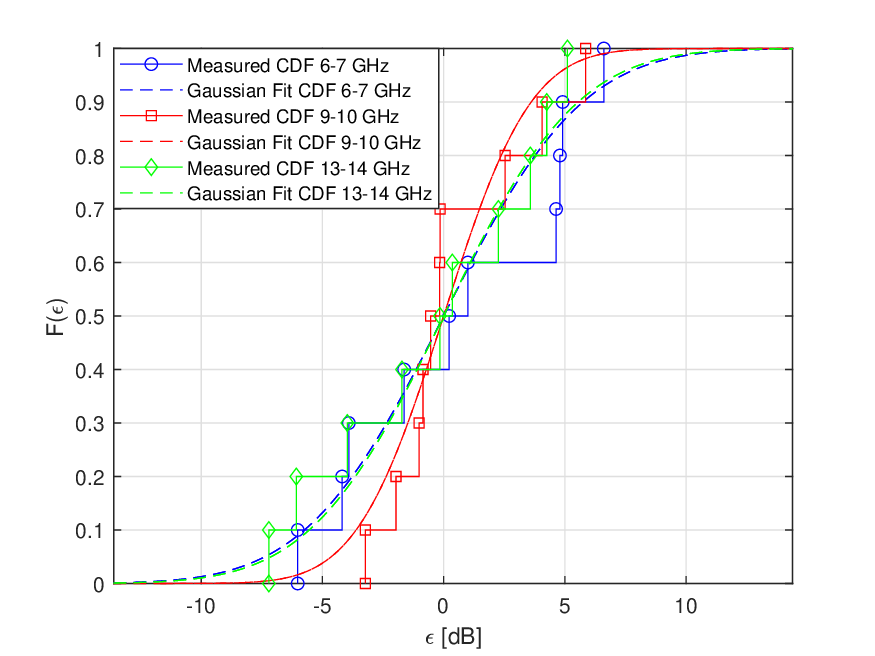}
        \caption{Shadowing.}
        \vspace*{0mm}
        \label{fig:PL_Omni2}
    \end{subfigure}
    \caption{\ac{OLoS} \ac{PL} modeling for Omni-directional case.}
    \label{fig:PL_Omni}
\end{figure*}

\begin{figure*}[t!]
\centering
    \begin{subfigure}{0.4\textwidth}
            \centering
            \hspace{7mm}
            \includegraphics[width=1\columnwidth]{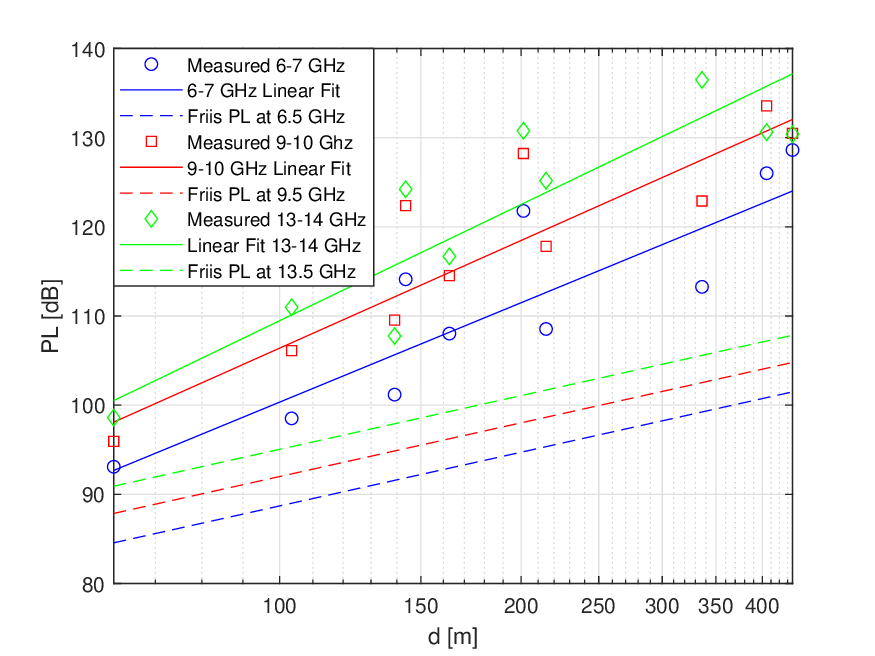}
            \caption{\ac{PL} modeling with $log_{10}(d)$ weighting.}
            \label{fig:PL_MD1}
    \end{subfigure}
    \begin{subfigure}{0.4\textwidth}
    \centering
        \centering
        \hspace{7mm}
        \includegraphics[width=1\columnwidth]{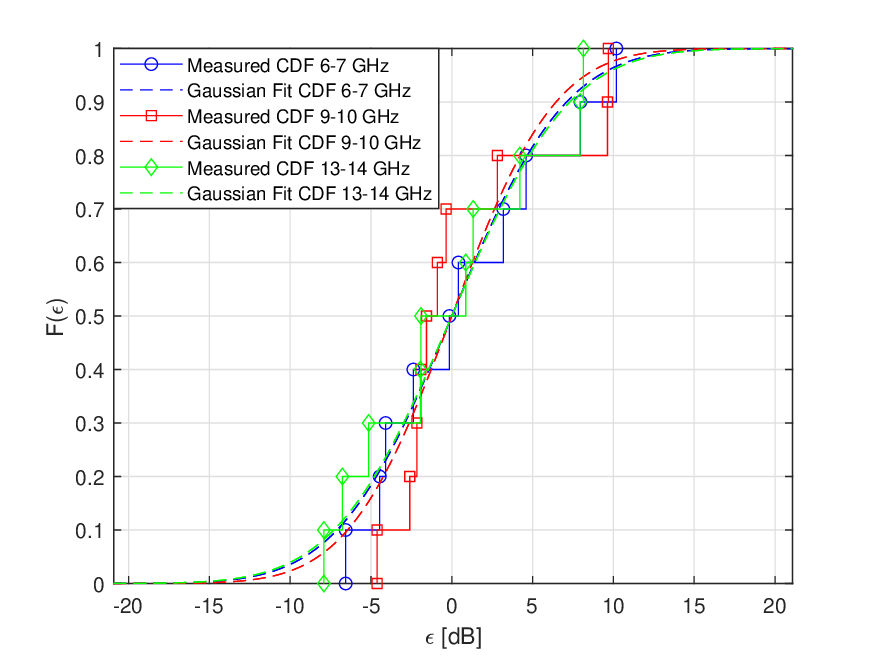}
        \caption{Shadowing.}
        \vspace*{0mm}
        \label{fig:PL_MD2}
    \end{subfigure}
    \caption{\ac{OLoS} \ac{PL} modeling for Max-Dir case.}
    \label{fig:PL_MD}
\end{figure*}

The path loss modeling results for the omni-directional case are shown in Fig. \ref{fig:PL_Omni1}, while the corresponding shadowing distribution is depicted in Fig. \ref{fig:PL_Omni2}. Similarly, Fig. \ref{fig:PL_MD1} and Fig. \ref{fig:PL_MD2} provide the same results for the Max-Dir case. The band-wise results for linear fitting and shadowing parameters are summarized in Tables \ref{tab:lf_PG_Omni}, \ref{tab:lf_PG_Max-Dir}, and \ref{tab:shadowing_params}. Please note that we remove the additional omni-directional gain for all different frequency bands.

\begin{table}[h!]
\centering
\caption{Linear fitting for $PL_{Omni}$ with 95$\%$ confidence intervals.}
\label{tab:lf_PG_Omni}
\begin{adjustbox}{width=8.5cm}
\begin{tabular}{|c||c|c|c|c|c|c|}
\hline
\textbf{Frequency} & $\boldsymbol{\alpha_{min,95\%}}$ & $\boldsymbol{\alpha}$ & $\boldsymbol{\alpha_{max,95\%}}$ & $\boldsymbol{\beta_{min,95\%}}$ & $\boldsymbol{\beta}$ & $\boldsymbol{\beta_{max,95\%}}$ \\ \hline \hline
All Bands & 3.26 & 21.76 & 40.25 & 3.32 & 4.14 & 4.97 \\ 
6-7 GHz & -5.40 & 22.25 & 49.91 & 2.64 & 3.87 & 5.10 \\ 
7-8 GHz & -0.55 & 24.23 & 49.01 & 2.78 & 3.88 & 4.98 \\ 
8-9 GHz & -4.37 & 19.18 & 42.72 & 3.08 & 4.13 & 5.17 \\ 
9-10 GHz & 2.67 & 19.67 & 36.68 & 3.49 & 4.25 & 5.00 \\ 
10-11 GHz & 4.40 & 20.13 & 35.87 & 3.61 & 4.31 & 5.01 \\ 
11-12 GHz & 7.28 & 23.56 & 39.84 & 3.45 & 4.17 & 4.89 \\ 
12-13 GHz & -11.23 & 16.27 & 43.77 & 3.28 & 4.51 & 5.73 \\ 
13-14 GHz & -10.65 & 15.36 & 41.38 & 3.41 & 4.57 & 5.72 \\ 
\hline
\end{tabular}
\end{adjustbox}
\end{table}

\begin{table}[h!]
\centering
\caption{Linear fitting for $PL_{\rm Max-Dir}$ with 95$\%$ confidence intervals.}
\label{tab:lf_PG_Max-Dir}
\begin{adjustbox}{width=8.5cm}
\begin{tabular}{|c||c|c|c|c|c|c|}
\hline
\textbf{Frequency} & $\boldsymbol{\alpha_{min,95\%}}$ & $\boldsymbol{\alpha}$ & $\boldsymbol{\alpha_{max,95\%}}$ & $\boldsymbol{\beta_{min,95\%}}$ & $\boldsymbol{\beta}$ & $\boldsymbol{\beta_{max,95\%}}$ \\ \hline \hline
All Bands & -2.98 & 28.57 & 60.11 & 2.50 & 3.90 & 5.30 \\ 
$6-7$ \SI{}{\giga\hertz} & -7.34 & 26.31 & 59.96 & 2.21 & 3.70 & 5.20 \\ 
$7-8$ \SI{}{\giga\hertz} & -3.38 & 30.12 & 63.63 & 2.15 & 3.64 & 5.13 \\ 
$8-9$ \SI{}{\giga\hertz} & -4.25 & 26.01 & 56.28 & 2.50 & 3.85 & 5.19 \\ 
$9-10$ \SI{}{\giga\hertz} & -3.61 & 26.20 & 56.01 & 2.68 & 4.01 & 5.33 \\ 
$10-11$ \SI{}{\giga\hertz} & -2.16 & 25.57 & 53.29 & 2.88 & 4.11 & 5.34 \\ 
$11-12$ \SI{}{\giga\hertz} & 2.58 & 31.00 & 59.41 & 2.63 & 3.90 & 5.16 \\ 
$12-13$ \SI{}{\giga\hertz} & -14.94 & 21.55 & 58.03 & 2.71 & 4.33 & 5.95 \\ 
$13-14$ \SI{}{\giga\hertz} & -9.98 & 22.89 & 55.76 & 2.87 & 4.33 & 5.79 \\ 
\hline
\end{tabular}
\end{adjustbox}
\end{table}

\begin{table}[h!]
\centering
\caption{Shadowing distribution parameters with 95$\%$ confidence intervals.}
\label{tab:shadowing_params}
\begin{adjustbox}{width=8.5cm}
\begin{tabular}{|c||c|c|c|c|c|c|}
\hline
\multirow{2}{*}{\textbf{Frequency}} & \multicolumn{3}{c|}{$\boldsymbol{PL_{Omni}}$} & \multicolumn{3}{c|}{$\boldsymbol{PL_{Max-Dir}}$} \\ \cline{2-7} 
& $\boldsymbol{\sigma}$ & $\boldsymbol{\sigma_{\text{min, 95\%}}}$ & $\boldsymbol{\sigma_{\text{max, 95\%}}}$ & $\boldsymbol{\sigma}$ & $\boldsymbol{\sigma_{\text{min, 95\%}}}$ & $\boldsymbol{\sigma_{\text{max, 95\%}}}$ \\ \hline \hline
All Bands & 3.22 & 1.36 & 4.05 & 5.46 & 1.76 & 6.93 \\ 
6-7 GHz & 4.48 & 2.76 & 5.35 & 5.54 & 3.06 & 6.79 \\ 
7-8 GHz & 4.22 & 2.38 & 5.48 & 5.77 & 3.31 & 7.34 \\ 
8-9 GHz & 3.99 & 2.22 & 5.01 & 5.20 & 2.04 & 6.62 \\ 
9-10 GHz & 2.82 & 1.09 & 3.57 & 5.03 & 1.50 & 6.25 \\ 
10-11 GHz & 2.60 & 1.50 & 3.11 & 4.74 & 1.66 & 6.17 \\ 
11-12 GHz & 2.70 & 1.26 & 3.32 & 4.89 & 1.41 & 6.15 \\ 
12-13 GHz & 4.67 & 2.52 & 5.49 & 6.22 & 3.16 & 7.44 \\ 
13-14 GHz & 4.33 & 2.57 & 5.29 & 5.69 & 3.56 & 7.00 \\ 
\hline
\end{tabular}
\end{adjustbox}
\end{table}

The \ac{PL} results reveal several key trends. First, path loss increases with frequency, as expected, due to the frequency-dependent nature of different types of attenuation. Similarly, path loss increases with distance, reflecting the fundamental behavior of signal attenuation over larger propagation ranges. However, the difference in \ac{PL} across various sub-bands slightly increases with frequency. We also note that while the parameters $\alpha$ and $\beta$ - representing the reference path loss and the path loss exponent, respectively - might not consistently increase over frequency, the path loss (computed over the range of observed distances) \textit{does} exhibit a clear increase with frequency. This highlights the dominant effect of frequency-dependent attenuation, even in the presence of environmental factors that are causing variability in the fitted parameters.

For pure free-space propagation, theory predicts a value of $\beta = 2$ independent of frequency, and an increase of $\alpha$ with $f^2$.
A comparison of the measured $\beta$ values with the theoretical value reveals interesting trends. For the omni-directional case (Table \ref{tab:lf_PG_Omni}), $\beta$ values range from $3.87$ ($6-7$\SI{}{\giga\hertz}) to $4.57$ ($13-14$ \SI{}{\giga\hertz}), significantly higher than the theoretical value. This deviation suggests that additional propagation factors, in particular vegetation loss, but also reflection and scattering, add a loss that increases with increasing distance, and that also changes the behavior with frequency. For the Max-Dir case (Table \ref{tab:lf_PG_Max-Dir}), $\beta$ values range from $3.64$ ($7-8$ \SI{}{\giga\hertz}) to $4.33$ ($13-14$ \SI{}{\giga\hertz}), which are slightly lower than the omni-directional values. This indicates that the frequency dependency of path loss is less pronounced in the Max-Dir case, likely due to the exclusion of some reflected \acp{MPC} that are captured in the omni-directional measurements.

The results also highlight the difference between omni-directional and Max-Dir cases. The latter exhibits lower path gain due to the stronger spatial filtering; note, however, that the gain of the \ac{Tx} and \ac{Rx} antennas will lead to a higher {\em link gain} than in the omni case. 


The Max-Dir case experiences slightly larger variance of the shadowing compared to the omni-directional case, Figs. \ref{fig:PL_Omni2} and \ref{fig:PL_MD2}, and Table V. This is likely due to the directional nature of Max-Dir measurements, which has a smaller number of strong paths, and thus less ``shadowing diversity" compared to the omnidirectional case. 

Wide confidence intervals are observed in certain sub-bands, particularly in the higher frequency ranges, as shown in Tables \ref{tab:lf_PG_Omni} and \ref{tab:lf_PG_Max-Dir}. 
These wider intervals suggest that the propagation characteristics at these frequencies are more variable, requiring more extensive data for robust parameter estimation.

\subsection{RMSDS}

\begin{figure*}[t!]
\centering
    \begin{subfigure}{0.4\textwidth}
    \centering
        \centering
        \hspace{7mm}
        \includegraphics[width=1\columnwidth]{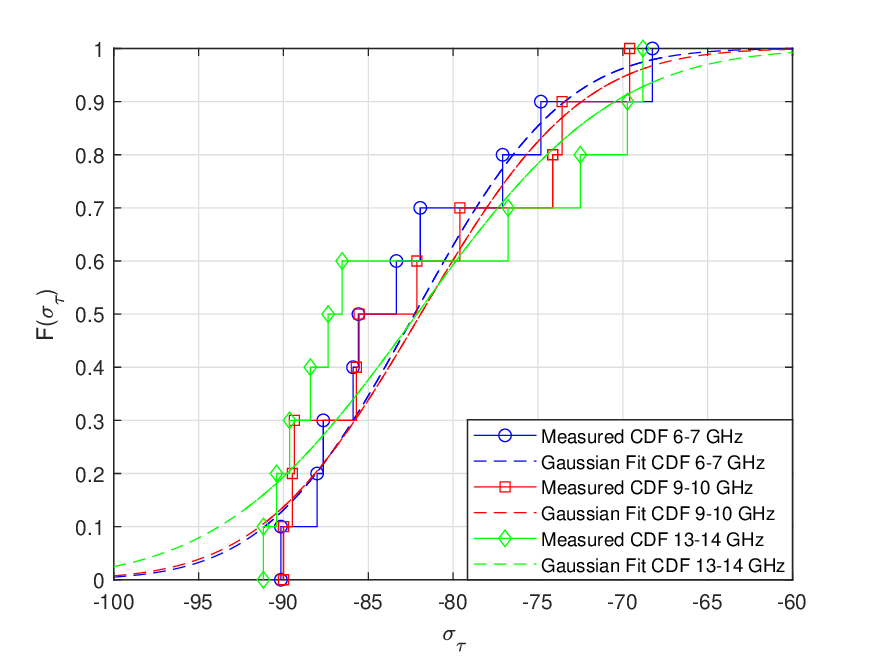}
        \caption{CDF of \ac{RMSDS}.}
        \vspace*{0mm}
        \label{fig:DS_Omni1}
    \end{subfigure}
    \begin{subfigure}{0.4\textwidth}
            \centering
            \hspace{7mm}
            \includegraphics[width=1\columnwidth]{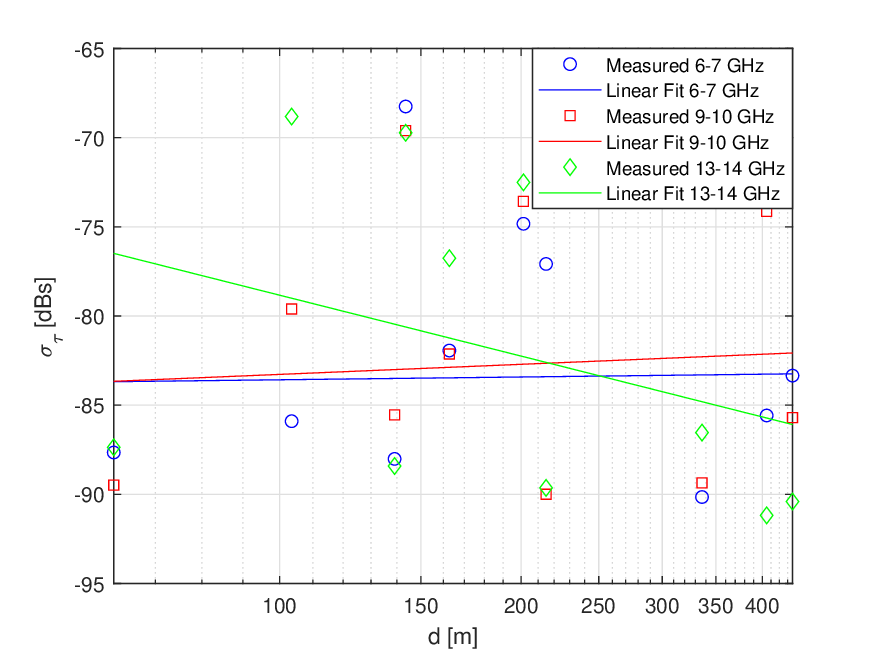}
            \caption{\ac{RMSDS} modeling with $log_{10}(d)$ weighting.}
            \label{fig:DS_Omni2}
    \end{subfigure}
    \caption{\ac{RMSDS} modeling for Omni-directional case.}
    \label{fig:DS_Omni}
\end{figure*}

\begin{figure*}[t!]
\centering
    \begin{subfigure}{0.4\textwidth}
    \centering
        \centering
        \hspace{7mm}
        \includegraphics[width=1\columnwidth]{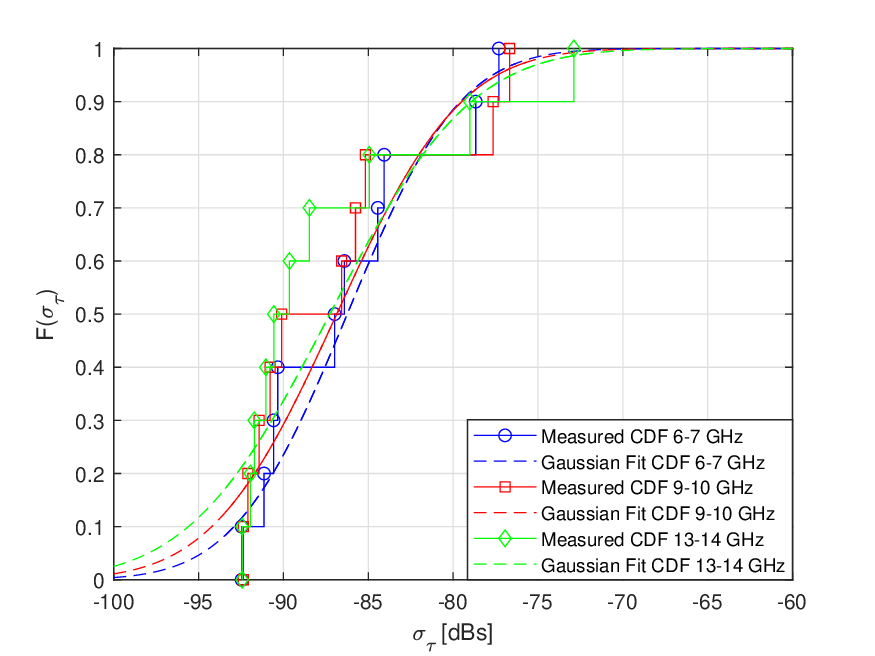}
        \caption{CDF of \ac{RMSDS}.}
        \vspace*{0mm}
        \label{fig:DS_MD1}
    \end{subfigure}
    \begin{subfigure}{0.4\textwidth}
            \centering
            \hspace{7mm}
            \includegraphics[width=1\columnwidth]{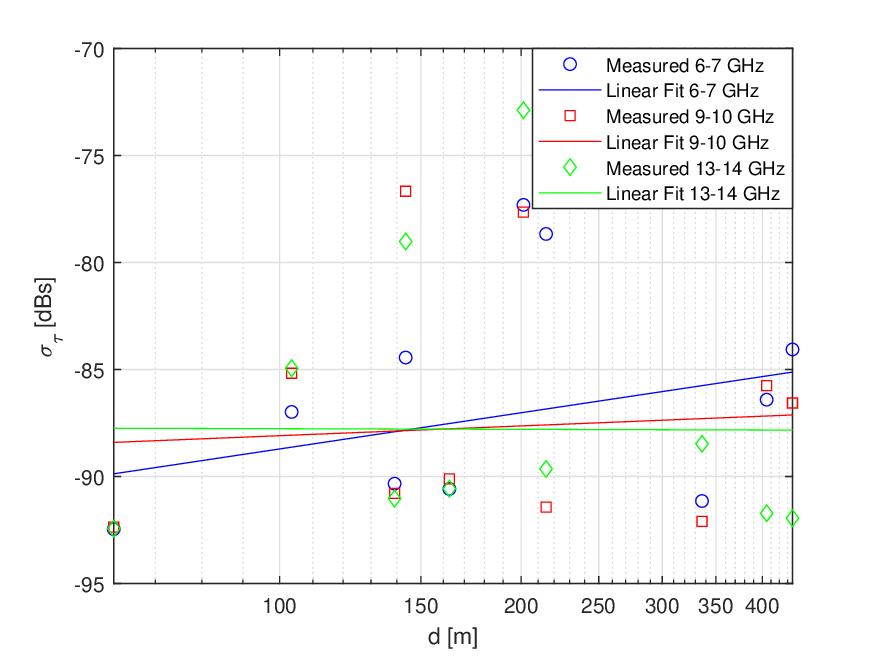}
            \caption{\ac{RMSDS} modeling with $log_{10}(d)$ weighting.}
            \label{fig:DS_MD2}
    \end{subfigure}
    \caption{\ac{RMSDS} modeling for Max-Dir case.}
    \label{fig:DS_MD}
\end{figure*}

The \acp{CDF} and corresponding fits of \ac{RMSDS} are shown in Fig. \ref{fig:DS_Omni1} for the omni-directional case and Fig. \ref{fig:DS_MD1} for the Max-Dir case. The corresponding modeling results as a function of $\log_{10}(d)$ are presented in Fig. \ref{fig:DS_Omni2} and Fig. \ref{fig:DS_MD2}. The bandwise fitting results are summarized in Tables \ref{tab:lf_RMDDS_Omni} (omni-directional) and \ref{tab:lf_RMSDS_Max-Dir} (Max-Dir), with the mean and standard deviation values provided in Tables \ref{tab:nf_RMDDS_Omni} (omni-directional) and \ref{tab:nf_RMSDS_Max-Dir} (Max-Dir).

\begin{table}[h!]
\centering
\caption{Linear fitting for $RMDDS_{\rm Omni}$ with 95$\%$ confidence intervals.}
\label{tab:lf_RMDDS_Omni}
\begin{adjustbox}{width=8.5cm}
\begin{tabular}{|c||c|c|c|c|c|c|}
\hline
\textbf{Frequency} & $\boldsymbol{\alpha_{min,95\%}}$ & $\boldsymbol{\alpha}$ & $\boldsymbol{\alpha_{max,95\%}}$ & $\boldsymbol{\beta_{min,95\%}}$ & $\boldsymbol{\beta}$ & $\boldsymbol{\beta_{max,95\%}}$ \\ \hline \hline
All Bands & -128.69 & -77.80 & -26.91 & -2.60 & -0.34 & 1.92 \\ 
$6-7$ \SI{}{\giga\hertz} & -126.36 & -84.61 & -42.87 & -1.80 & 0.05 & 1.91 \\ 
$7-8$ \SI{}{\giga\hertz} & -140.92 & -89.41 & -37.90 & -2.03 & 0.26 & 2.55 \\ 
$8-9$ \SI{}{\giga\hertz} & -122.93 & -71.27 & -19.60 & -2.82 & -0.53 & 1.77 \\ 
$9-10$ \SI{}{\giga\hertz} & -132.01 & -87.03 & -42.04 & -1.81 & 0.19 & 2.19 \\ 
$10-11$ \SI{}{\giga\hertz} & -125.23 & -84.51 & -43.79 & -1.75 & 0.06 & 1.87 \\ 
$11-12$ \SI{}{\giga\hertz} & -116.54 & -72.45 & -28.36 & -2.31 & -0.35 & 1.61 \\ 
$12-13$ \SI{}{\giga\hertz} & -126.02 & -78.75 & -31.48 & -2.33 & -0.22 & 1.88 \\ 
$13-14$ \SI{}{\giga\hertz} & -110.40 & -56.16 & -1.91 & -3.54 & -1.13 & 1.28 \\ 
\hline
\end{tabular}
\end{adjustbox}
\end{table}

\begin{table}[h!]
\centering
\caption{Normal fitting for $RMDDS_{\rm Omni}$ with 95$\%$ confidence intervals.}
\label{tab:nf_RMDDS_Omni}
\begin{adjustbox}{width=8.5cm}
\begin{tabular}{|l||c|c|c|c|c|c|}
\hline
\textbf{Frequency} & $\boldsymbol{\mu_{min,95\%}}$ & $\boldsymbol{\mu}$ & $\boldsymbol{\mu_{max,95\%}}$ & $\boldsymbol{\sigma_{min,95\%}}$ & $\boldsymbol{\sigma}$ & $\boldsymbol{\sigma_{max,95\%}}$ \\ \hline \hline
All Bands & -90.68 & -84.54 & -78.39 & 5.91 & 8.59 & 15.69 \\ 
$6-7$ \SI{}{\giga\hertz} & -87.21 & -82.27 & -77.34 & 4.74 & 6.90 & 12.59 \\ 
$7-8$ \SI{}{\giga\hertz} & -88.60 & -82.57 & -76.54 & 5.80 & 8.43 & 15.38 \\ 
$8-9$ \SI{}{\giga\hertz} & -87.87 & -82.32 & -76.77 & 5.34 & 7.76 & 14.17 \\ 
$9-10$ \SI{}{\giga\hertz} & -87.20 & -81.91 & -76.62 & 5.09 & 7.40 & 13.51 \\ 
$10-11$ \SI{}{\giga\hertz} & -87.03 & -82.29 & -77.55 & 4.56 & 6.63 & 12.10 \\ 
$11-12$ \SI{}{\giga\hertz} & -85.68 & -80.56 & -75.44 & 4.93 & 7.16 & 13.07 \\ 
$12-13$ \SI{}{\giga\hertz} & -88.29 & -82.71 & -77.12 & 5.37 & 7.80 & 14.25 \\ 
$13-14$ \SI{}{\giga\hertz} & -88.65 & -82.13 & -75.62 & 6.26 & 9.10 & 16.62 \\ 
\hline
\end{tabular}
\end{adjustbox}
\end{table}

\begin{table}[h!]
\centering
\caption{Linear fitting for $RMSDS_{\rm Max-Dir}$ with 95$\%$ confidence intervals.}
\label{tab:lf_RMSDS_Max-Dir}
\begin{adjustbox}{width=8.5cm}
\begin{tabular}{|c||c|c|c|c|c|c|}
\hline
\textbf{Frequency} & $\boldsymbol{\alpha_{min,95\%}}$ & $\boldsymbol{\alpha}$ & $\boldsymbol{\alpha_{max,95\%}}$ & $\boldsymbol{\beta_{min,95\%}}$ & $\boldsymbol{\beta}$ & $\boldsymbol{\beta_{max,95\%}}$ \\ \hline \hline
All Bands & -153.33 & -111.54 & -69.76 & -0.93 & 0.93 & 2.78 \\ 
$6-7$ \SI{}{\giga\hertz} & -130.59 & -99.95 & -69.30 & -0.80 & 0.56 & 1.92 \\ 
$7-8$ \SI{}{\giga\hertz} & -123.82 & -93.56 & -63.30 & -1.10 & 0.24 & 1.59 \\ 
$8-9$ \SI{}{\giga\hertz} & -123.28 & -96.57 & -69.86 & -0.83 & 0.36 & 1.55 \\ 
$9-10$ \SI{}{\giga\hertz} & -125.89 & -91.12 & -56.35 & -1.39 & 0.15 & 1.70 \\ 
$10-11$ \SI{}{\giga\hertz} & -131.48 & -101.41 & -71.34 & -0.74 & 0.60 & 1.94 \\ 
$11-12$ \SI{}{\giga\hertz} & -121.87 & -90.54 & -59.20 & -1.26 & 0.13 & 1.53 \\ 
$12-13$ \SI{}{\giga\hertz} & -137.65 & -91.91 & -46.18 & -1.85 & 0.18 & 2.22 \\ 
$13-14$ \SI{}{\giga\hertz} & -125.42 & -87.56 & -49.71 & -1.69 & -0.01 & 1.67 \\ 
\hline
\end{tabular}
\end{adjustbox}
\end{table}

\begin{table}[h!]
\centering
\caption{Normal fitting for $RMSDS_{\rm Max-Dir}$ with 95$\%$ confidence intervals.}
\label{tab:nf_RMSDS_Max-Dir}
\begin{adjustbox}{width=8.5cm}
\begin{tabular}{|l||c|c|c|c|c|c|}
\hline
\textbf{Frequency} & $\boldsymbol{\mu_{min,95\%}}$ & $\boldsymbol{\mu}$ & $\boldsymbol{\mu_{max,95\%}}$ & $\boldsymbol{\sigma_{min,95\%}}$ & $\boldsymbol{\sigma}$ & $\boldsymbol{\sigma_{max,95\%}}$ \\ \hline \hline
All Bands & -94.64 & -89.77 & -84.91 & 4.68 & 6.80 & 12.42 \\ 
$6-7$ \SI{}{\giga\hertz} & -89.97 & -86.24 & -82.51 & 3.58 & 5.21 & 9.51 \\ 
$7-8$ \SI{}{\giga\hertz} & -91.18 & -87.50 & -83.81 & 3.54 & 5.15 & 9.41 \\ 
$8-9$ \SI{}{\giga\hertz} & -91.12 & -87.82 & -84.52 & 3.17 & 4.61 & 8.42 \\ 
$9-10$ \SI{}{\giga\hertz} & -90.97 & -86.86 & -82.75 & 3.95 & 5.75 & 10.50 \\ 
$10-11$ \SI{}{\giga\hertz} & -90.99 & -87.11 & -83.24 & 3.73 & 5.42 & 9.89 \\ 
$11-12$ \SI{}{\giga\hertz} & -90.56 & -86.94 & -83.32 & 3.48 & 5.06 & 9.24 \\ 
$12-13$ \SI{}{\giga\hertz} & -92.33 & -86.68 & -81.02 & 5.43 & 7.90 & 14.42 \\ 
$13-14$ \SI{}{\giga\hertz} & -91.92 & -87.27 & -82.61 & 4.48 & 6.51 & 11.88 \\ 
\hline
\end{tabular}
\end{adjustbox}
\end{table}

The results show that \ac{RMSDS} varies significantly with respect to distance, but no clear increasing or decreasing trend is observed for either the omni-directional or Max-Dir cases. The modeling of \ac{RMSDS} versus distance produces wide confidence intervals, particularly for higher frequency sub-bands. For instance, Table \ref{tab:lf_RMDDS_Omni} shows that the parameter $\beta$, which indicates the change of \ac{RMSDS} with $\log_{10}(d)$, ranges from $-3.54$ to $1.92$ \SI{}{\decibel\second} ($13-14$ \SI{}{\giga\hertz} and $6-7$ \SI{}{\giga\hertz}, respectively), crossing zero in several sub-bands. This suggests that \ac{RMSDS} does not strongly correlate with distance in the current measurement environment, likely due to the effects of scattering and multipath being highly variable across measurement locations. For example, the strong, long-delayed component at Rx5 previously discussed, is the result of the particular arrangement of buildings at this location, providing a much larger delay spread than at most locations with either shorter or longer distance to the \ac{Tx}.  

Despite these variations with distance, the values of \ac{RMSDS} distributions appear relatively stable across different frequency bands, as indicated by the mean values in Tables \ref{tab:nf_RMDDS_Omni} and \ref{tab:nf_RMSDS_Max-Dir}. For the omni-directional case, the mean \ac{RMSDS} across all bands is approximately $-84.54$ \SI{}{\decibel\second} with a standard deviation of $8.59$ \SI{}{\decibel\second} (Table \ref{tab:nf_RMDDS_Omni}), while for the Max-Dir case, the mean is approximately $-89.77$ \SI{}{\decibel\second} with a standard deviation of $6.80$ \SI{}{\decibel\second} (Table \ref{tab:nf_RMSDS_Max-Dir}). Examining band-by-band values, the omni-directional \ac{RMSDS} ranges from $-82.27$ \SI{}{\decibel\second} ($6-7$ \SI{}{\giga\hertz}) to $-82.13$ \SI{}{\decibel\second} ($13-14$ \SI{}{\giga\hertz}) (Table \ref{tab:nf_RMDDS_Omni}), while the Max-Dir values range from $-86.24$ \SI{}{\decibel\second} ($6-7$ \SI{}{\giga\hertz}) to $-87.27$ \SI{}{\decibel\second} ($13-14$ \SI{}{\giga\hertz}) (Table \ref{tab:nf_RMSDS_Max-Dir}). These small changes across bands indicate that the delay spread does not vary significantly with frequency for either the omni-directional or Max-Dir cases.

This stability suggests that \ac{RMSDS} is not strongly frequency-dependent in the current urban microcellular scenario. Furthermore, the relatively consistent \ac{RMSDS} values across frequency bands indicate that the aggregate multipath propagation characteristics remain stable, regardless of variations in dynamic range or environmental conditions, though individual MPCs might still show stronger variations across frequency (this will be investigated in future work). 

\subsection{Angular spread}

\begin{figure*}[t!]
\centering
    \begin{subfigure}{0.4\textwidth}
    \centering
        \centering
        \hspace{7mm}
        \includegraphics[width=1\columnwidth]{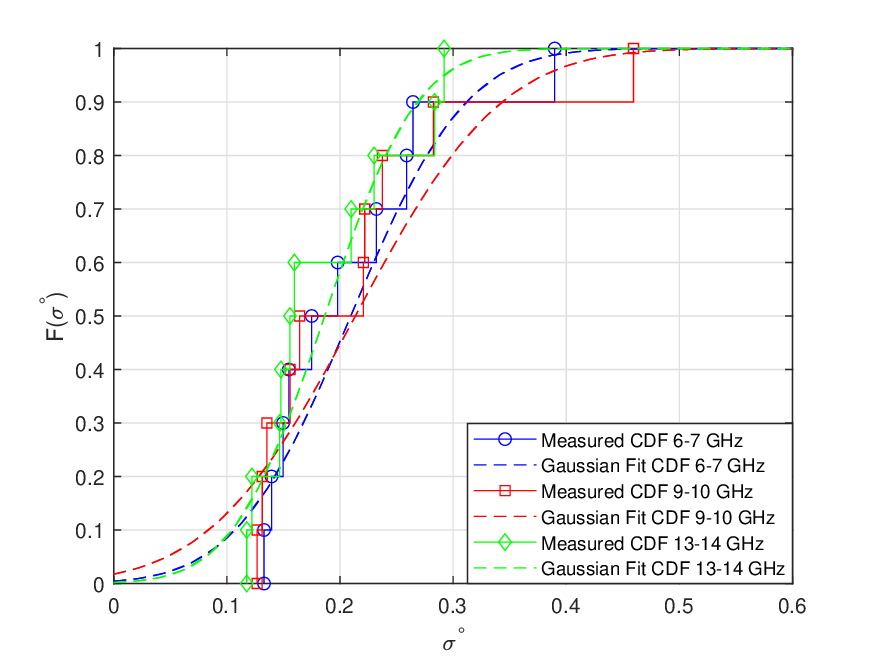}
        \caption{CDF of \ac{AS}.}
        \vspace*{0mm}
        \label{fig:AS_TX1}
    \end{subfigure}
    \begin{subfigure}{0.4\textwidth}
            \centering
            \hspace{7mm}
            \includegraphics[width=1\columnwidth]{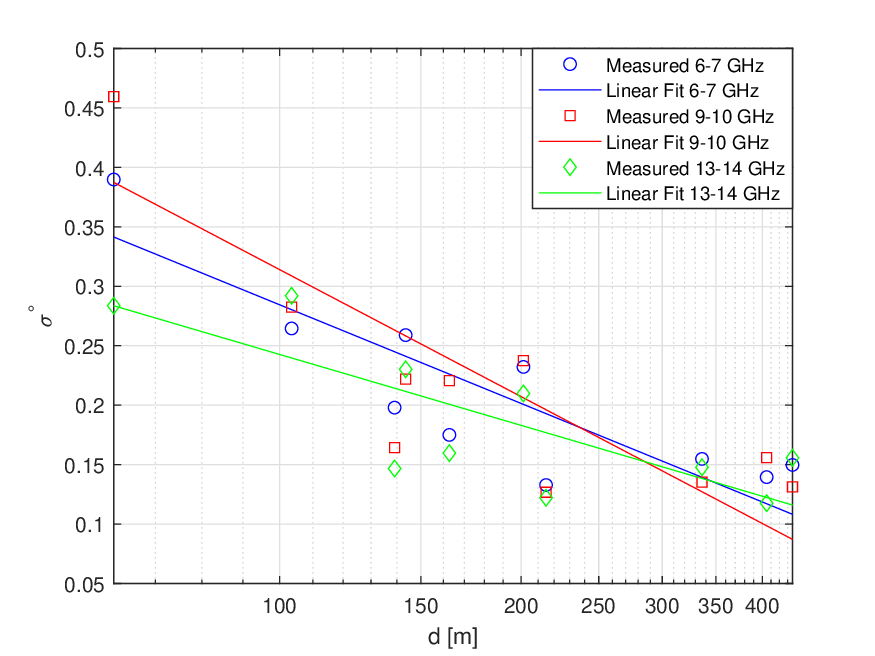}
            \caption{\ac{AS} modeling with $log_{10}(d)$ weighting.}
            \label{fig:AS_TX2}
    \end{subfigure}
    \caption{\ac{AS} modeling for \ac{Tx}.}
    \label{fig:AS_TX}
\end{figure*}

\begin{figure*}[t!]
\centering
    \begin{subfigure}{0.4\textwidth}
    \centering
        \centering
        \hspace{7mm}
        \includegraphics[width=1\columnwidth]{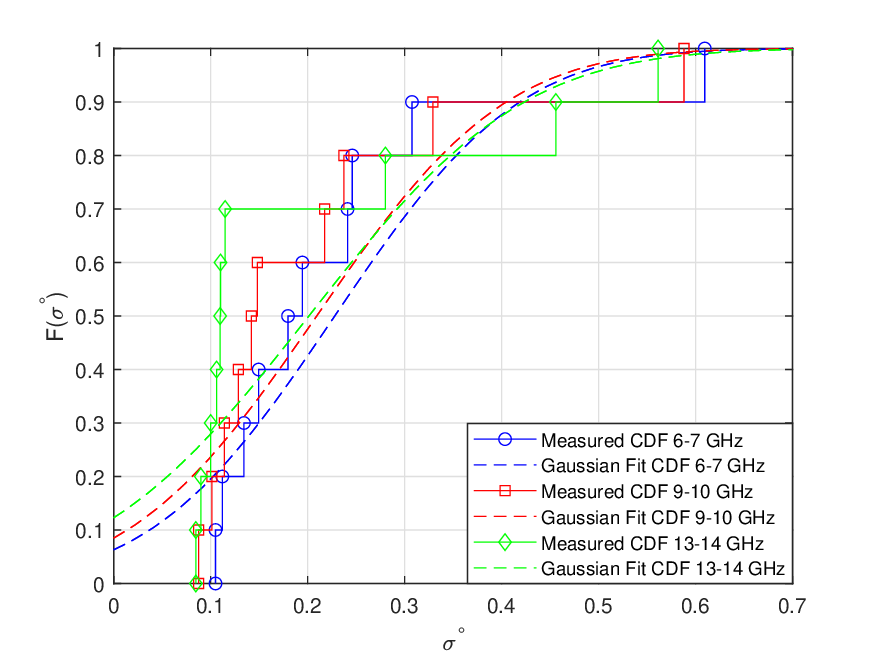}
        \caption{CDF of \ac{AS}.}
        \vspace*{0mm}
        \label{fig:AS_RX1}
    \end{subfigure}       
    \begin{subfigure}{0.4\textwidth}
            \centering
            \hspace{7mm}
            \includegraphics[width=1\columnwidth]{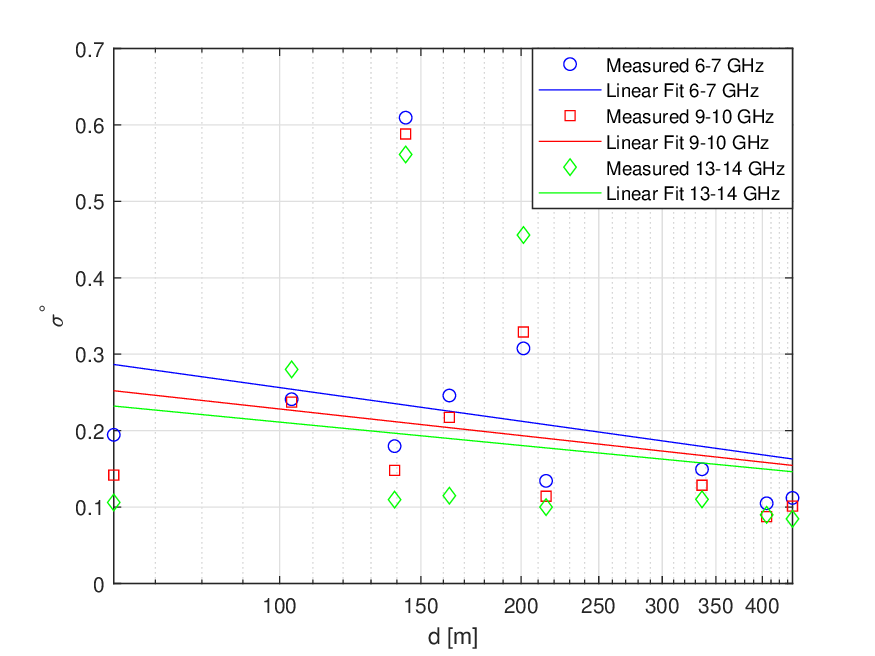}
            \caption{\ac{AS} modeling with $log_{10}(d)$ weighting.}
            \label{fig:AS_RX2}  
    \end{subfigure}    
    \caption{\ac{AS} modeling for \ac{Rx}.}
    \label{fig:AS_TX}
\end{figure*}

\ac{AS} provides insights into the spatial characteristics of the channel at both the \ac{Tx} and \ac{Rx}. The \ac{CDF}s of \ac{AS} are presented in Fig. \ref{fig:AS_TX1} for the \ac{Tx} and Fig. \ref{fig:AS_RX1} for the \ac{Rx}, while the corresponding modeling as a function of $\log_{10}(d)$ is shown in Fig. \ref{fig:AS_TX2} and Fig. \ref{fig:AS_RX2}. The results, summarized in Tables \ref{tab:lf_AS_TX}, \ref{tab:lf_AS_RX-AZ}, and \ref{tab:lf_AS_RX-El}, demonstrate trends across different frequency bands, ranging from 6 \si{\giga\hertz} to 14 \si{\giga\hertz}. Additionally, the mean and standard deviation values, provided in Tables \ref{tab:nf_AS_TX}, \ref{tab:nf_AS_RX-AZ}, and \ref{tab:nf_AS_RX-El}, highlight the statistical variation of \ac{AS} across the measured scenarios.

\begin{table}[h!]
\centering
\caption{Linear fitting for $AS_{\rm TX}$ with 95$\%$ confidence intervals.}
\label{tab:lf_AS_TX}
\begin{adjustbox}{width=8.5cm}
\begin{tabular}{|c||c|c|c|c|c|c|}
\hline
\textbf{Frequency} & $\boldsymbol{\alpha_{min,95\%}}$ & $\boldsymbol{\alpha}$ & $\boldsymbol{\alpha_{max,95\%}}$ & $\boldsymbol{\beta_{min,95\%}}$ & $\boldsymbol{\beta}$ & $\boldsymbol{\beta_{max,95\%}}$ \\ \hline \hline
All Bands & 0.60 & 0.92 & 1.23 & -0.05 & -0.03 & -0.02 \\ 
$6-7$ \SI{}{\giga\hertz} & 0.57 & 0.84 & 1.10 & -0.04 & -0.03 & -0.02 \\ 
$7-8$ \SI{}{\giga\hertz} & 0.67 & 1.01 & 1.36 & -0.05 & -0.03 & -0.02 \\ 
$8-9$ \SI{}{\giga\hertz} & 0.57 & 0.94 & 1.31 & -0.05 & -0.03 & -0.02 \\ 
$9-10$ \SI{}{\giga\hertz} & 0.68 & 1.02 & 1.37 & -0.05 & -0.04 & -0.02 \\ 
$10-11$ \SI{}{\giga\hertz} & 0.50 & 0.74 & 0.99 & -0.03 & -0.02 & -0.01 \\ 
$11-12$ \SI{}{\giga\hertz} & 0.63 & 0.92 & 1.20 & -0.04 & -0.03 & -0.02 \\ 
$12-13$ \SI{}{\giga\hertz} & 0.51 & 0.79 & 1.06 & -0.04 & -0.03 & -0.01 \\ 
$13-14$ \SI{}{\giga\hertz} & 0.38 & 0.64 & 0.89 & -0.03 & -0.02 & -0.01 \\ 
\hline
\end{tabular}
\end{adjustbox}
\end{table}

\begin{table}[h!]
\centering
\caption{Normal fitting for $AS_{\rm TX}$ with 95$\%$ confidence intervals.}
\label{tab:nf_AS_TX}
\begin{adjustbox}{width=8.5cm}
\begin{tabular}{|l||c|c|c|c|c|c|}
\hline
\textbf{Frequency} & $\boldsymbol{\mu_{min,95\%}}$ & $\boldsymbol{\mu}$ & $\boldsymbol{\mu_{max,95\%}}$ & $\boldsymbol{\sigma_{min,95\%}}$ & $\boldsymbol{\sigma}$ & $\boldsymbol{\sigma_{max,95\%}}$ \\ \hline \hline
All Bands & 0.14 & 0.20 & 0.27 & 0.06 & 0.09 & 0.16 \\ 
$6-7$ \SI{}{\giga\hertz} & 0.15 & 0.21 & 0.27 & 0.05 & 0.08 & 0.15 \\ 
$7-8$ \SI{}{\giga\hertz} & 0.14 & 0.21 & 0.29 & 0.07 & 0.10 & 0.18 \\ 
$8-9$ \SI{}{\giga\hertz} & 0.14 & 0.20 & 0.27 & 0.07 & 0.09 & 0.17 \\ 
$9-10$ \SI{}{\giga\hertz} & 0.14 & 0.21 & 0.29 & 0.07 & 0.10 & 0.18 \\ 
$10-11$ \SI{}{\giga\hertz} & 0.15 & 0.20 & 0.25 & 0.05 & 0.07 & 0.13 \\ 
$11-12$ \SI{}{\giga\hertz} & 0.13 & 0.20 & 0.26 & 0.06 & 0.09 & 0.16 \\ 
$12-13$ \SI{}{\giga\hertz} & 0.13 & 0.19 & 0.24 & 0.05 & 0.08 & 0.14 \\ 
$13-14$ \SI{}{\giga\hertz} & 0.14 & 0.19 & 0.23 & 0.04 & 0.06 & 0.12 \\ 
\hline
\end{tabular}
\end{adjustbox}
\end{table}

\begin{table}[h!]
\centering
\caption{Linear fitting for $AS_{\rm RX-AZ}$ with 95$\%$ confidence intervals.}
\label{tab:lf_AS_RX-AZ}
\begin{adjustbox}{width=8.5cm}
\begin{tabular}{|c||c|c|c|c|c|c|}
\hline
\textbf{Frequency} & $\boldsymbol{\alpha_{min,95\%}}$ & $\boldsymbol{\alpha}$ & $\boldsymbol{\alpha_{max,95\%}}$ & $\boldsymbol{\beta_{min,95\%}}$ & $\boldsymbol{\beta}$ & $\boldsymbol{\beta_{max,95\%}}$ \\ \hline \hline
All Bands & -0.33 & 0.45 & 1.22 & -0.05 & -0.01 & 0.02 \\ 
$6-7$ \SI{}{\giga\hertz} & -0.25 & 0.55 & 1.34 & -0.05 & -0.01 & 0.02 \\ 
$7-8$ \SI{}{\giga\hertz} & -0.28 & 0.43 & 1.14 & -0.04 & -0.01 & 0.02 \\ 
$8-9$ \SI{}{\giga\hertz} & -0.14 & 0.41 & 0.96 & -0.03 & -0.01 & 0.01 \\ 
$9-10$ \SI{}{\giga\hertz} & -0.38 & 0.46 & 1.30 & -0.05 & -0.01 & 0.03 \\ 
$10-11$ \SI{}{\giga\hertz} & -0.53 & 0.43 & 1.38 & -0.05 & -0.01 & 0.03 \\ 
$11-12$ \SI{}{\giga\hertz} & -0.57 & 0.47 & 1.52 & -0.06 & -0.01 & 0.03 \\ 
$12-13$ \SI{}{\giga\hertz} & -0.81 & 0.37 & 1.55 & -0.06 & -0.01 & 0.04 \\ 
$13-14$ \SI{}{\giga\hertz} & -0.58 & 0.41 & 1.40 & -0.05 & -0.01 & 0.03 \\ 
\hline
\end{tabular}
\end{adjustbox}
\end{table}

\begin{table}[h!]
\centering
\caption{Normal fitting for $AS_{\rm RX-AZ}$ with 95$\%$ confidence intervals.}
\label{tab:nf_AS_RX-AZ}
\begin{adjustbox}{width=8.5cm}
\begin{tabular}{|l||c|c|c|c|c|c|}
\hline
\textbf{Frequency} & $\boldsymbol{\mu_{min,95\%}}$ & $\boldsymbol{\mu}$ & $\boldsymbol{\mu_{max,95\%}}$ & $\boldsymbol{\sigma_{min,95\%}}$ & $\boldsymbol{\sigma}$ & $\boldsymbol{\sigma_{max,95\%}}$ \\ \hline \hline
All Bands & 0.10 & 0.20 & 0.31 & 0.10 & 0.14 & 0.26 \\ 
$6-7$ \SI{}{\giga\hertz} & 0.12 & 0.23 & 0.33 & 0.10 & 0.15 & 0.27 \\ 
$7-8$ \SI{}{\giga\hertz} & 0.11 & 0.20 & 0.30 & 0.09 & 0.13 & 0.24 \\ 
$8-9$ \SI{}{\giga\hertz} & 0.11 & 0.19 & 0.26 & 0.07 & 0.10 & 0.18 \\ 
$9-10$ \SI{}{\giga\hertz} & 0.10 & 0.21 & 0.32 & 0.10 & 0.15 & 0.28 \\ 
$10-11$ \SI{}{\giga\hertz} & 0.09 & 0.21 & 0.34 & 0.12 & 0.17 & 0.32 \\ 
$11-12$ \SI{}{\giga\hertz} & 0.09 & 0.22 & 0.36 & 0.13 & 0.19 & 0.34 \\ 
$12-13$ \SI{}{\giga\hertz} & 0.07 & 0.22 & 0.37 & 0.14 & 0.21 & 0.38 \\ 
$13-14$ \SI{}{\giga\hertz} & 0.08 & 0.20 & 0.33 & 0.12 & 0.17 & 0.32 \\ 
\hline
\end{tabular}
\end{adjustbox}
\end{table}

\begin{table}[h!]
\centering
\caption{Linear fitting for $AS_{\rm RX-El}$ with 95$\%$ confidence intervals.}
\label{tab:lf_AS_RX-El}
\begin{adjustbox}{width=8.5cm}
\begin{tabular}{|c||c|c|c|c|c|c|}
\hline
\textbf{Frequency} & $\boldsymbol{\alpha_{min,95\%}}$ & $\boldsymbol{\alpha}$ & $\boldsymbol{\alpha_{max,95\%}}$ & $\boldsymbol{\beta_{min,95\%}}$ & $\boldsymbol{\beta}$ & $\boldsymbol{\beta_{max,95\%}}$ \\ \hline \hline
All Bands & 0.12 & 0.25 & 0.39 & -0.01 & -0.01 & -0.00 \\ 
$6-7$ \SI{}{\giga\hertz} & 0.14 & 0.27 & 0.39 & -0.01 & -0.01 & -0.00 \\ 
$7-8$ \SI{}{\giga\hertz} & 0.13 & 0.26 & 0.38 & -0.01 & -0.01 & -0.00 \\ 
$8-9$ \SI{}{\giga\hertz} & 0.13 & 0.29 & 0.44 & -0.01 & -0.01 & -0.00 \\ 
$9-10$ \SI{}{\giga\hertz} & 0.11 & 0.27 & 0.42 & -0.01 & -0.01 & 0.00 \\ 
$10-11$ \SI{}{\giga\hertz} & 0.10 & 0.27 & 0.43 & -0.01 & -0.01 & 0.00 \\ 
$11-12$ \SI{}{\giga\hertz} & 0.08 & 0.22 & 0.35 & -0.01 & -0.00 & 0.00 \\ 
$12-13$ \SI{}{\giga\hertz} & 0.06 & 0.24 & 0.43 & -0.01 & -0.01 & 0.00 \\ 
$13-14$ \SI{}{\giga\hertz} & 0.11 & 0.24 & 0.37 & -0.01 & -0.01 & 0.00 \\ 
\hline
\end{tabular}
\end{adjustbox}
\end{table}

\begin{table}[h!]
\centering
\caption{Normal fitting for $AS_{\rm RX-El}$ with 95$\%$ confidence intervals.}
\label{tab:nf_AS_RX-El}
\begin{adjustbox}{width=8.5cm}
\begin{tabular}{|l||c|c|c|c|c|c|}
\hline
\textbf{Frequency} & $\boldsymbol{\mu_{min,95\%}}$ & $\boldsymbol{\mu}$ & $\boldsymbol{\mu_{max,95\%}}$ & $\boldsymbol{\sigma_{min,95\%}}$ & $\boldsymbol{\sigma}$ & $\boldsymbol{\sigma_{max,95\%}}$ \\ \hline \hline
All Bands & 0.09 & 0.11 & 0.13 & 0.02 & 0.03 & 0.05 \\ 
$6-7$ \SI{}{\giga\hertz} & 0.10 & 0.12 & 0.14 & 0.02 & 0.03 & 0.05 \\ 
$7-8$ \SI{}{\giga\hertz} & 0.09 & 0.11 & 0.14 & 0.02 & 0.03 & 0.05 \\ 
$8-9$ \SI{}{\giga\hertz} & 0.09 & 0.11 & 0.14 & 0.02 & 0.03 & 0.06 \\ 
$9-10$ \SI{}{\giga\hertz} & 0.10 & 0.12 & 0.14 & 0.02 & 0.03 & 0.06 \\ 
$10-11$ \SI{}{\giga\hertz} & 0.08 & 0.11 & 0.13 & 0.02 & 0.03 & 0.06 \\ 
$11-12$ \SI{}{\giga\hertz} & 0.09 & 0.11 & 0.13 & 0.02 & 0.03 & 0.05 \\ 
$12-13$ \SI{}{\giga\hertz} & 0.08 & 0.10 & 0.13 & 0.03 & 0.04 & 0.07 \\ 
$13-14$ \SI{}{\giga\hertz} & 0.09 & 0.11 & 0.13 & 0.02 & 0.03 & 0.05 \\ 
\hline
\end{tabular}
\end{adjustbox}
\end{table}

The \ac{AS} for the \ac{Tx} shows a clear dependence on distance, with the angular spread decreasing as the distance between the \ac{Tx} and \ac{Rx} increases. This behavior is consistent with the expectation that the effective angular dispersion reduces at larger distances due to the narrowing of the spatial channel. However, as shown in Fig. \ref{fig:AS_TX2}, the decrease is modest, with $\beta$ values ranging from $-0.05$ to $-0.02$ across the frequency bands (Table \ref{tab:lf_AS_TX}). These small negative values suggest a weak dependence on distance, and the \ac{AS} remains relatively stable for most measurement points.  

The \ac{Rx} angular spread, on the other hand, exhibits slightly larger values compared to the \ac{Tx} due to the wider azimuthal angular range covered by the \ac{Rx} ($360^\circ$) vs \ac{Tx} ($120^\circ$). This trend is evident in Fig. \ref{fig:AS_RX1}, where the \ac{CDF} with values up to $0.6$ for the \ac{Rx} \ac{AS} in comparison to Fig. \ref{fig:AS_TX1}, where the maximum values are below $0.5$. Fig. \ref{fig:AS_RX2} also show a slight decrease in \ac{AS} with distance; however, some data points deviate significantly from the linear fit, resulting in wide confidence intervals, particularly at higher frequencies. For instance, Table \ref{tab:lf_AS_RX-AZ} shows $\beta$ values ranging from $-0.06$ to $0.02$, with some intervals containing zero, indicating minimal or inconsistent dependence on distance in certain sub-bands.  

Interestingly, the \ac{AS} distributions for both the \ac{Tx} and \ac{Rx} across different frequency bands are quite similar, with only slight reductions in mean values as frequency increases. For example, the mean \ac{AS} for the \ac{Tx} varies from $0.21$ to $0.19$ (Table \ref{tab:nf_AS_TX}), while for the \ac{Rx}, the mean varies from $0.23$ to $0.20$ across the same frequency range (Table \ref{tab:nf_AS_RX-AZ}). 

The angular spread in elevation for the \ac{Rx} is relatively small compared to the azimuthal angular spread due to the limited range of elevation angles measured during the campaign. Specifically, the measurements were conducted across co-elevations spanning only $-20^\circ$ to $20^\circ$ at the \ac{Rx}, while the \ac{Tx} remained fixed at a single elevation. This restricted range naturally results in lower values for elevation spread. The modeling results (Table \ref{tab:lf_AS_RX-El}) show minimal variation with distance, with $\beta$ values close to zero, further emphasizing the constrained angular range in the vertical plane. Additionally, the distributions across frequency bands (Table \ref{tab:nf_AS_RX-El}) indicate limited variation in the mean and standard deviation, suggesting that elevation spread remains consistent regardless of distance or frequency in this urban microcellular environment.

\section{Conclusions}

This paper reports the first \ac{UWB} double-directionally resolved measurement campaign in the upper midband (FR 3) for urban microcells. Measurements are performed mainly in a street canyon with \ac{LoS} obstructed by vegetation. From the results, we can draw the following main conclusions:

\begin{itemize}
    \item Considering the \ac{PDP}s, the presence of obstructions, such as vegetation and buildings, along with multipath reflections, results in significant deviations from idealized free-space propagation. The \ac{LoS} case (Rx1) experiences received power above the predictions of the Friis model due to contributions from multiple \ac{MPC}s, while the \ac{OLoS} case (Rx5) falls significantly below Friis predictions due to vegetation attenuation and path obstructions.  

    \item Distinct differences are observed between the \ac{LoS} and \ac{OLoS} scenarios in the \ac{PDP} as well as the \ac{APS}. The \ac{LoS} case is dominated by a strong direct component, whereas the \ac{OLoS} case features more balanced contributions from multiple reflection clusters in comparison to the \ac{LoS} component.  

    \item The analysis of \ac{RMSDS} reveals relatively stable mean delay spread values across frequency bands, while standard deviations increase modestly with frequency. This is in contrast to assumptions in 3GPP that postulate a decrease of the mean with frequency. Dependence of the \ac{RMSDS} on distance is weak or inconsistent. 

    \item The angular spread analysis highlights expected differences between the \ac{Tx} and \ac{Rx} due to their spatial coverage areas. At the \ac{Tx}, angular spread shows some dependence on distance, particularly in the street canyon environment, while the variation across frequency bands is minimal. The \ac{Rx} angular spread is generally larger due to its $360^\circ$ azimuthal coverage, though the elevation angular spread is small, potentially due to the limited measured elevation range.

    \item The current campaign involved over 25,000 measured transfer functions, yet the number of unique \ac{Tx}-\ac{Rx} location pairs remains somewhat limited (10 \ac{OLoS} points and 1 \ac{LoS} point). This initial channel model offers realistic insights for system design, but future campaigns with more locations will increase the robustness and generality of the results. 
    
    \item The measurement locations were chosen to ensure reasonable \ac{Rx} power, conditioning the analysis on locations that support viable communication links. A ``blind" selection of points on a regular grid would provide a broader assessment, including outage probabilities and insights into regions where communication may not be sustainable, however, such a (future) study would require a time-domain sounder with a front-end based on electronically switched arrays in order to perform the resulting large number of measurements in realistic time.

    \item Despite its limitations, this study represents the first double-directional \ac{UWB} measurement campaign conducted in the \ac{OLoS} scenarios in the upper-midband for outdoor environments. The time-intensive nature of such measurements underscores the need for further refinement in future studies.  
\end{itemize}


\section*{Acknowledgment}
Helpful discussions and support from Xinzhou Su, Zihang Cheng, and Yuning Zhang are gratefully acknowledged.


\end{document}